\documentclass[%
reprint,
twocolumn,
superscriptaddress,
 amsmath,amssymb,
 aps,
]{revtex4-2}

\usepackage{graphicx}
\usepackage{bm}
\usepackage{xcolor}
\usepackage{amsfonts}
\usepackage{amssymb}
\usepackage{amsmath}
\usepackage{hyperref}
\usepackage{dsfont}
\usepackage[utf8]{inputenc}
\usepackage{bbm}

\usepackage{tikz}
\usetikzlibrary{quantikz2}
\usepackage{algorithm}
\usepackage{algpseudocode}

\begin{document}

\newcommand{\HR}[1] {\textcolor{blue}{({\bf HR:} #1)}}
\newcommand{\HRe}{\textcolor{blue}}
\newcommand{\NK}[1] {\textcolor{purple}{({\bf NK:} #1)}}
\newcommand{\JS}{\textcolor{red}}
\newcommand{\TB}{\textcolor{red}}
\newcommand{\SKR}[1]{\textcolor[HTML]{FF0F99}{#1}}
\newcommand{\MS}[1] {\textcolor{orange}{({\bf MS:} #1)}}

\preprint{APS/123-QED}

\title{Complex fluid flow dynamics by bright scattershot boson sampling}

\author{Robert Schade}
\affiliation{Paderborn Center for Parallel Computing, Paderborn University,
D-33098 Paderborn, Germany.}

\author{Shailendra K. Rathor}
\affiliation{Institute of Thermodynamics and Fluid Mechanics, Technische Universität Ilmenau, D-98684 Ilmenau, Germany}

\author{Tim J. Bartley}
\affiliation{Institute for Photonic Quantum Systems (PhoQS), Paderborn University, D-33098 Paderborn, Germany}
\affiliation{Department of Physics, Paderborn University, D-33098 Paderborn, Germany}

\author{Benjamin Brecht}
\affiliation{Institute for Photonic Quantum Systems (PhoQS), Paderborn University, D-33098 Paderborn, Germany}
\affiliation{Department of Physics, Paderborn University, D-33098 Paderborn, Germany}

\author{Christian Plessl}
\affiliation{Paderborn Center for Parallel Computing, Paderborn University,
D-33098 Paderborn, Germany.}
\affiliation{Department of Computer Science, Paderborn University,
D-33098 Paderborn, Germany.}

\author{Jan Sperling}
\affiliation{Institute for Photonic Quantum Systems (PhoQS), Paderborn University, D-33098 Paderborn, Germany}
\affiliation{Department of Physics, Paderborn University, D-33098 Paderborn, Germany}

\author{Michael Stefszky}
\affiliation{Institute for Photonic Quantum Systems (PhoQS), Paderborn University, D-33098 Paderborn, Germany}
\affiliation{Department of Physics, Paderborn University, D-33098 Paderborn, Germany}

\author{Christine Silberhorn}
\affiliation{Institute for Photonic Quantum Systems (PhoQS), Paderborn University, D-33098 Paderborn, Germany}
\affiliation{Department of Physics, Paderborn University, D-33098 Paderborn, Germany}

\author{Jörg Schumacher}
\affiliation{Institute of Thermodynamics and Fluid Mechanics, Technische Universität Ilmenau, D-98684 Ilmenau, Germany}

\date{\today}

\begin{abstract}
Boson sampling machines have become a fruitful tool for establishing new concepts, which connect fundamental physics with computer science applications. Here, we exploit the intrinsically high-dimensional correlations generated by the measurement outcomes of a multiphoton boson sampling machine and demonstrate how a reconfigurable architecture can be programmed to autonomously model the time evolution of nonlinear dynamical systems that describe complex fluid flows. By extending on the original scattershot boson sampling (SBS) scheme to include higher photon number components -- a scheme which we label bright scattershot boson sampling (BSBS) --- this scheme overcomes scaling issues inherent to SBS-based schemes by utilizing all randomly generated input photon number distributions. We illustrate the capability of this scheme for two complex fluid-flow models: an eight-dimensional Lorenz-type model of two-dimensional thermally driven convection and a 16-dimensional model of global atmospheric circulation. Using a digital twin of the BSBS architecture based on multiphoton statistics, we show that the photonic approach accurately reproduces the temporal dynamics of both systems while being resilient to photon losses of up to 50\%. We also verified that the BSBS case outperforms a comparable run with coherent light.  Moreover, its prediction horizon exceeds that of a classical reservoir-computing model with the same query complexity. These results highlight the potential of near-term photonic quantum computing for addressing relevant real-world tasks on noisy intermediate-scale devices.
\end{abstract}

\maketitle

\section{Introduction \label{sec:introduction}}

Boson sampling \cite{Aaronson2011} and its extension to Gaussian input states \cite{Lund2014,Hamilton2017,Kruse2019} has emerged as a route to demonstrating quantum computational advantage with photonic systems  \cite{Wang2019,zhong2020,Madsen2022,Romero2024,Fu2026}. While initially regarded as an abstract concept, albeit one with important implications connecting fundamental principles of physics and computer science, the solving of tasks with practical application has since been demonstrated, including the modeling of molecular spectra \cite{Huh2015,Zhu2024} and clique detection in macromolecular networks \cite{Yu2023}. Initially, these tasks were motivated by the possibility of demonstrating a practical quantum advantage, however, follow-up investigations led to the development of more efficient classical solutions \cite{Oh2024,Lim2025} and better understanding of where the complexity in these problems lies \cite{Eickmann2026}. These findings motivated the search for investigations into alternative applications, such as Monte Carlo integration  \cite{Andersen2025,Andersen2025est,anguita2025} and asset clustering \cite{lopena2026}, as well as alternative boson sampling architectures \cite{Spagnolo2023,Hoch2025}, where it is hoped that a practical quantum advantage may be found. 


An alternative avenue for realizing practical quantum advantage is in using these boson sampling machines to implement (quantum) machine learning tasks \cite{rambach2026,Paparelle2026,Yin2025,Sakurai2022,Hoch2025,Cimini2026,Zhu2025,gong2026,Sharifian2026,Pfeffer2022,Henaff24,GarciaBeni2025}. In particular, boson sampling-based machines have been primarily used for classification and image recognition~\cite{rambach2026,gong2026,Sharifian2026,Sakurai2022,Cimini2026}. Quantum machine learning (QML) algorithms successfully exploit many-particle quantum systems to perform classification, regression, and prediction tasks, as well as generate new synthetic data \cite{Biamonte2017,Schuld2019,Cerezo2022,Rudolph2022}. 
QML is one of the most promising applications of Boson samplers because these systems provide a natively large Hilbert space. Furthermore, some system biases are expected to be compensated for by the algorithms, see e.g. a discussion in \cite{Herteux2020} in classical cases. 
In this mode of operation, the boson sampler can be regarded as a ``quantum extreme learning machine,'' or more generally a quantum reservoir computing model (RCM) \cite{Jaeger2004,Lukosevicius2009,Tanaka2019}.

Recently, first proof-of-principle demonstrations of reservoir computing have been realised on photonic platforms. Nerenberg et al. \cite{Nerenberg2025} proposed a photon-number-resolving quantum RCM and discussed several input schemes based on different polarisations. Paparelle et al. \cite{Paparelle2026} used squeezed states to forecast one-dimensional nonlinear time series. Selimovi\'{c} et al. \cite{Selimovic2026} presented a reservoir computing model based on a photonic quantum memristor implemented using a tunable Mach-Zehnder interferometer. They applied this model to predict one-dimensional nonlinear time series and the dynamics of a single degree of freedom governed by the chaotic Mackey-Glass equation \cite{Mackey1977}. More recently, a boson-sampling architecture has been used to demonstrate time-series forecasting and investigate the advantage offered by indistinguishable photons \cite{DiBartolo2026,Fu2026}. Building on these proof-of-principle demonstrations, the field is now exploring schemes that enable scaling in the number of input states, system size, or feature-space dimension. Further applications beyond classification are also being sought to address real-world problems, with particular interest in schemes capable of investigating higher-dimensional nonlinear systems.

In this work, we present a scheme that addresses scalability by exploiting the high dimensionality of the Fock space of bright parametric down-conversion sources as a deterministic generator of two-mode squeezed vacuum (TMSV) states. In this way, we purposely exploit higher-order photon-pair emission events to expand the available parameter space and thus efficiently harness higher-order correlation functions. The presented scheme is therefore an extension to what is commonly referred to as scattershot boson sampling, which we coin bright scattershot boson sampling (BSBS). We also introduce an encoding scheme to predict complex fluid-flow dynamics without solving the underlying nonlinear partial differential equations. For training the system the full multiport interferometer is programmed such that the scheme investigates trajectories in higher-dimensional state spaces than those considered in previous approaches \cite{DiBartolo2026}, taking a step towards modelling fully developed turbulence on a quantum photonic platform. To this end, we map a reservoir computing model (RCM) onto a reconfigurable bright scattershot boson sampler, exploiting the intrinsically high-dimensional correlations among measurement outcomes.

We evaluate this approach using a ``digital twin'' of a small-scale multiphoton boson sampler, simulated on a high-performance computer. Using this simulated device, we predict the evolution of two nonlinear dynamical systems with up to 16 degrees of freedom that capture complex flow behavior. We benchmark its performance against a classical RCM and estimate the device scale beyond which simulating the photonic model and reconstructing the dynamics would exceed the computational resources available on our high-performance computer. For the more complex learning task, the photonic RCM achieves a prediction horizon twice that of its classical counterpart. These results outline a route towards practically relevant dynamical prediction using scattershot boson sampling, with an implementation within reach of existing quantum photonic technology.

\section{Multiphoton reservoir computing model}

\begin{figure*}[t!]
	\centering
	\includegraphics[width = \textwidth]{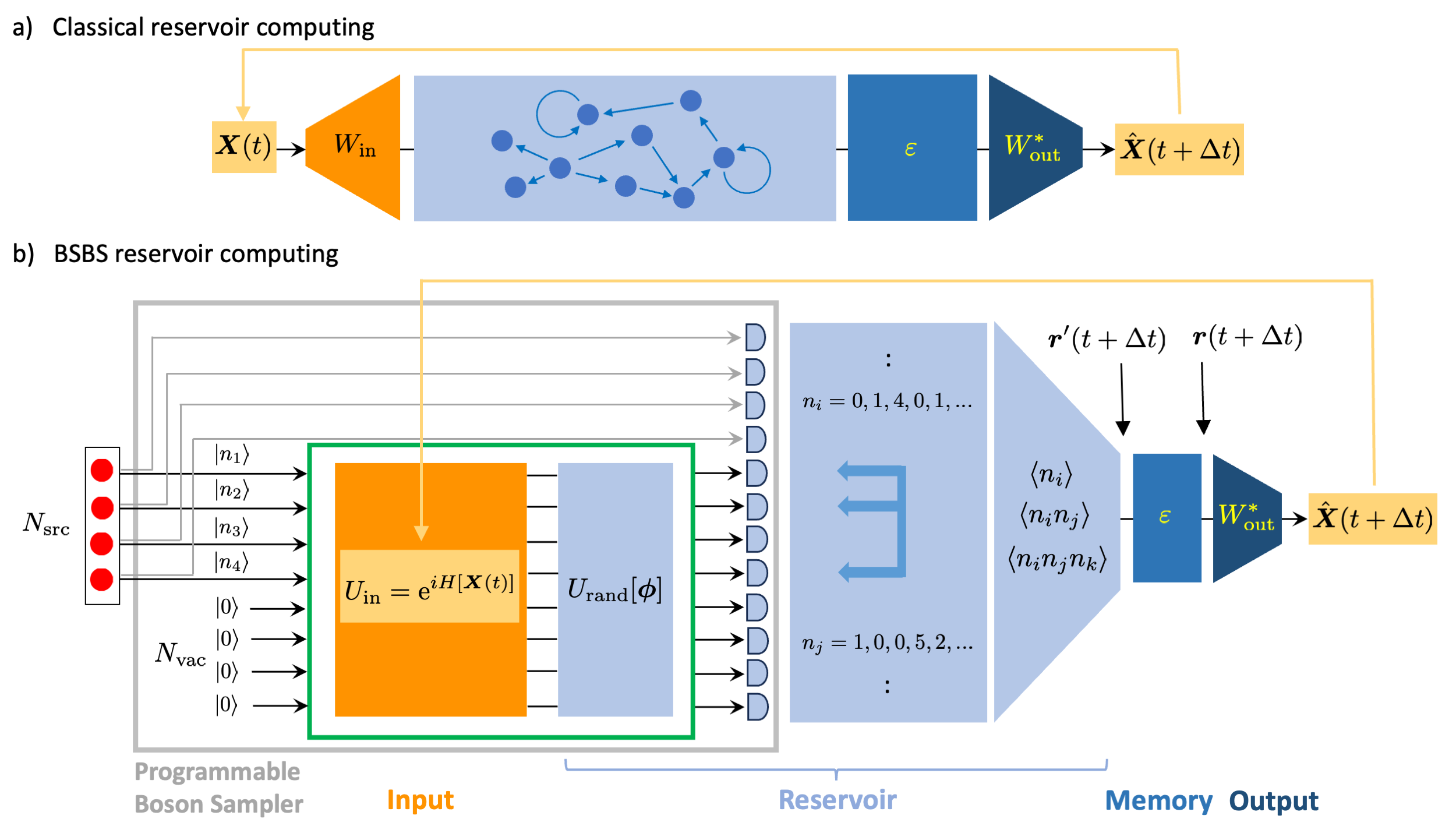}
	\caption{{\bf Closed-loop reservoir computing models.} The output data vector of the reservoir computing model at time $t+\Delta t$ is the programmable input for the next time step. Both models consist of an input layer $W_{\rm in}$ (orange), a reservoir with memory (light blue) and an optimized output layer $W_{\rm out}^*$ (dark blue). a) Classical reservoir computing model (C-RCM). The reservoir is a sparse network of neurons. b) Bright scattershot boson sampling (BSBS) reservoir computing model. Spontaneous parametric down-conversion photon sources create $N_{\rm src}$ TMSV states. One mode of each TMSV state is directly detected which heralds the number of photons in the second mode of the TMSV that enter different channels of our linear network. The input unitary $U_{\rm in}$ encodes the dynamical system state ${\bm X}(t)$ at time step $t$ (here the past reservoir output) via a self-adjoint Hamiltonian. In combination with a random unitary $U_{\rm rand}[{\bm \phi}]$ a linear optical network $U_{\rm total}=U_{\rm rand}[{\bm \phi}] U_{\rm in}$ of phase shifters and beam splitters processes the input. The green box therefore represents the programmable interferometer, which implements $U_{\rm total}$. The photon number detection at network exit, which includes all measured detection events, generates a high-dimensional photon number distribution with probabilities. A subsequent reduction layer is inserted to downscale the high-dimensional GBS output vector by taking photon number correlations at the $2N_{\rm src}+N_{\rm vac}$ modes and determine photon number moments $\langle n_i\rangle$, $\langle n_i n_j\rangle$, and in some cases even $\langle n_i n_j n_k\rangle$. See also Table~\ref{tab:runs} for details. The reduced data vector ${\bm r}'(t+\Delta t)$ enters the optimized linear output layer $W^*_{\rm out}$ of the reservoir after blending with the past state ${\bm r}(t)$ by leaking rate $\varepsilon$. This leads to the dynamical system prediction $\hat{\bm X}(t+\Delta t)$ which is the input for the next processing step.}
	\label{fig:Fig1}
\end{figure*}
\subsection{Building blocks}
Photonic reservoir computing models predict dynamics with quantum states of light \cite{Mujal2021,Labay2024}. Figure \ref{fig:Fig1} compares classical and photonic RCMs; both consist of three building blocks, an input layer, the reservoir with memory, and an output layer, the latter of which is optimized only in a training procedure. Next to the one-step open-loop scenario, RCMs can be set up in a {\em closed-loop regime}, which will be considered here, see Fig.~\ref{fig:Fig1}. In this regime, the predicted output state $\hat{\bm X}(t)\in \mathbb{R}^{N_{\rm dof}}$ is the input ${\bm X}(t)\in \mathbb{R}^{N_{\rm dof}}$ to obtain $\hat{\bm X}(t+\Delta t)$ in the next iteration step of the RCM. Here, $N_{\rm dof}$ is the number of degrees of freedom.  

A limiting case of reservoir computing models are extreme learning machines, algorithms that have no short-term memory and consist of a single hidden layer in the classical configuration \cite{Huang2004}. In an extreme learning machine, the quantum reservoir state ${\bm \Psi}$ at time $t+\Delta t$ is solely determined by the classical data input at time $t$. This class includes nonlinear vector autoregression algorithm, also termed next generation reservoir computing \cite{Gauthier2021,Jaurigue2022}. Qubit-based implementations of extreme learning machines have been discussed in \cite{Xiong2023,Innocenti2023},  and a small scale proof-of-principle demonstration with photonic systems in ref. \cite{Wang2025}. 

Classical and quantum reservoir computing models ${\cal R}$ follow an update rule with short-term memory for the quantum reservoir state, which is given by
\begin{equation}
{\bm \Psi}(t+\Delta t) = {\cal R}[{\bm \Psi}(t), {\bm X}(t)]\,.
\end{equation}
This enables us to model nonlinear chaotic dynamics such as turbulence, that is always characterized by finite-time correlations \cite{Fujii2017,Chen2020,Mujal2021,Araiza2022,Pfeffer2022,Pfeffer2023,Ahmed2024,Kobayashi2024}. Figure~\ref{fig:Fig1}~(a) shows a classical reservoir computing model, contrasted with our quantum RCM shown in panel (b). In our model, the time protocol is a so-called feedback-driven protocol \cite{Pfeffer2022,Pfeffer2023,Kobayashi2024}. A measurement at time step $t$ determines the next step at $t+\Delta t$; the state of the quantum reservoir is destroyed (since we apply a projective measurement), but information from these measurements is fed back into the quantum RCM cycle. Other time protocols are discussed in \cite{Fujii2017,Mujal2023,Cindrak2024}. 

We design our approach to be compatible with the Paderborn Quantum Sampler, an academic Boson Sampling machine \cite{stefszky2025}. The pipeline of the data processing in the quantum photonic experiment is illustrated in  Fig.~\ref{fig:Fig1}(b). The steps from photon generation, encoding of inputs in the interferometer, photon detection, reduction to correlation functions, reservoir step, and finally prediction of the dynamical quantities in the next time step are described in the following sections step by step.

\subsection{Photon generation}
The starting point is $N_{\rm src}$ photon sources that generate single spectro-temporal TMSV states through spontaneous parametric down conversion (SPDC) \cite{luo2026}, indicated as red spots in Fig.~\ref{fig:Fig1}(b). The SPDC process non-deterministically converts one or more photons of high energy into one or more pairs of photons of lower energy, known as signal and idler; measuring one mode of the downconverted fields heralds the presence of photons in the other field, which enters a programmable interferometer implementing a unitary transformation. Each state is given by
\begin{equation}
\ket{\psi}_\textrm{SPDC}=\sqrt{1-\lambda^2}\sum_{n=0}^\infty\lambda^n\ket{n,n}~,
\label{photonsource}
\end{equation}
which describes the creation of $n$ photons in each of two modes, parameterised by $0\leq\lambda<1$. This is related to the mean photon number generated by the source $\bar{n}=\lambda^2/(1-\lambda^2)$. It has been shown that one can achieve mean photon numbers of up to ${\cal O}(10^4)$ \cite{luo2026}, corresponding to an upper bound $\lambda>0.9995$. The joint photon number distribution $P(n)$ of a certain photon number being spontaneously emitted in each mode is $P(n, \lambda)=\left(1-\lambda^2\right)\lambda^{2n}$.      

\subsection{Encoding of inputs}
The photons enter the programmable interferometer as a Fock state $|\Psi_{\rm in}\rangle =|n_1,n_2,...,n_{N_{\rm src}},0,0,...,0\rangle$ into which the data vector ${\bm X}$ is encoded. The unitary matrix $U_{\rm total}$ of the interferometer is formed as a product of a randomly chosen unitary matrix $U_{\rm rand}[{\bm \phi}]$, which can be seen as part of the reservoir, and an encoding unitary matrix $U_{\rm in}[{\bm X}(t)]$ as
\begin{equation}
U_{\rm total}(t)=U_{\rm rand}[{\bm \phi}] U_{\rm in}[{\bm X}(t)] \in U(N_{\rm src}+N_{\rm vac}).
\end{equation}
For each time step $t$, a different interferometer matrix $U_{\rm total}(t)$ is programmed into the interferometer after performing the measurement, i.e., a reconfigurable optical chip comprising a linear optics network composed of beam splitters and phase shifters where angles and splitting ratios at the phase shifters and beam splitters, respectively can be chosen to implement the given unitary matrix, see Fig.~\ref{fig:Fig1}(b). 
The random reservoir-matrix can be parametrized by a real vector ${\bm \phi} \in \mathbb{R}^{(N_{\rm src}+N_{\rm vac})^2}$ which is kept fixed. The inputs ${\bm X}(t)$ are encoded into the unitary encoding matrix $U_{\rm in}(t)$ via a Hermitian matrix $H$ as
\begin{equation}
U_{\rm in}(t)=\exp(iH[{\bm X}(t)])
\end{equation}
with the dynamical variables on the off-diagonals of the Hermitian matrix, e.g. for four dynamical inputs $\bm X(t)=(X_1(t),X_2(t),X_3(t),X_4(t))$ in a four-dimensional unitary
\begin{equation}
 H[{\bm X}(t)]=\mathbbm{1}_{4}+\alpha \begin{pmatrix}
    0 & X_1(t) & X_2(t) & X_3(t) \\
    X_1(t) & 0 & X_4(t) & 0 & \\
    X_2(t) & X_4(t) & 0 & 0 &  \\
    X_3(t) & 0 & 0 & 0
\end{pmatrix}.
\end{equation}
The parameter $\alpha$ can be used to adjust to the dynamic range of the dynamical variables and is kept fixed. Throughout this work, we have used $\alpha=0.01$. Even though the encoding matrix $U_{\rm in}$ has a rather regular structure due to the underlying structure of the Hermitian matrix $H$, the product with the random reservoir matrix $U_{\rm rand}[{\bm \phi}]$ will result in an unstructured and fully occupied interferometer matrix $U_{\rm total}$ with high probability.

\subsection{Photon detection and sampling}
The generated photons pass through the programmed interferometer, resulting in the state $|\Psi_{\rm out}\rangle$, as
\begin{equation}
|\Psi_{\rm out}\rangle = \hat U_{\rm total}(t) |\Psi_{\rm in}\rangle\, ,
\end{equation}
where $\hat U_{\rm total}(t)$ denotes the many-particle operator that corresponds to the action of the interferometer with the unitary matrix $U_{\rm total}(t)$.
Subsequent photon-number resolving detection results in a Fock state $|\Psi_{\rm out,i}\rangle=|n_{i,1},n_{i,2},...,n_{i,2N_{\rm src}+N_{\rm vac}}\rangle$ with $i$ being sample number index. Repeating this sampling procedure $N_{\rm sample}$ times results in a set of samples $\{n_{i,1},n_{i,2},...\}_{i=1,...,N_{\rm sample}}$ of photon counts of the $2N_{\rm src}+N_{\rm vac}$ modes. For $N_{\rm sample}\rightarrow \infty$, these samples represent the photon number distributions in the state $|\Psi_{\rm out}\rangle$.

\subsection{Reduction to mode-correlation functions}
The measured photon number probability distribution, i.e., the measurement probabilities of the states $|n_1,n_2,...,n_{2N_{\rm src+}+N_{\rm vac}}\rangle$ for the individual interferometer configuration $ U_{\rm total}(t)$ could now be used in various ways as the inputs for the reservoir memory and linear output layer. However, using the state probabilities directly introduces an optimal selection problem where states would have to be selected in the photon number space such that their probability can, on the one hand, be accurately measured with a low number of samples, and on the other hand is sufficiently dynamically dependent on the dynamical quantities $\bm X(t)$ via the interferometer unitary $U_{\rm total}(t)$. Even though experimentally available detectors are limited to distinguishing $n_{{\rm max}}$ photons at each detector, for example 3 or 4 photons in the case of superconducting nanowire single-photon detectors \cite{Schapeler2024}, the available photon number space grows exponentially with the number of interferometer modes. To avoid this selection problem, instead the samples are reduced to first-, second- and higher-order moments of the particle number, i.e., the expectation values 
\begin{align}
    \langle \hat n_i \hat n_j ...\rangle&=\sum_{a=1}^{N_{\rm sample}} \langle n_{a,1},n_{a,2},... | \hat n_i \hat n_j ... | n_{a,1},n_{a,2}, ... \rangle \nonumber\\
    &=\sum_{a=1}^{N_{\rm sample}} n_{a,i}n_{a,j}... \label{eq:photonmoments}
\end{align}
or, in other words, photon number expectation values $\langle \hat n_i\rangle$ and multi-mode correlation functions $\langle \hat n_i\hat n_j ...\rangle$.
The maximum photon-number moment $O_m$, i.e, maximal number of terms in the products $ n_{a,i}n_{a,j}...$ determines the total number of expectation values $N_{\rm obs}$, which is given by 
\begin{equation}
N_{\rm obs}=\sum_{k=1}^{O_m} \frac{(\tilde{N}+k-1)!}{k!(\tilde{N}-1)!}\;\mbox{with}\;
\tilde{N}=2N_{\rm src}+N_{\rm vac}
\end{equation}
for $\tilde{N}=2N_{\rm src}+N_{\rm vac}$ modes because photon number operators commute. Hence, the number of expectation values grows like $\mathcal{O}(\tilde{N}^{O_m})$ for $O_m\ll \tilde{N}$ instead of $\mathcal{O}((n_{{\rm max}}+1)^{\tilde{N}})$ if using the state probabilities. In this work, $O_m=2$ and $O_m=3$ are examined, and Table~\ref{tab:runs} lists the corresponding counts for the investigated examples.

\begin{table}[ht]
\centering
\begin{tabular}{lccccc}
\hline\hline
Label & $N_{\rm mod}$ & $N_{\rm src}$ & $N_{\rm vac}$ & $O_m$ & $N_{\rm obs}$ \\
\hline
L8a / L96a & 8  & 4 & 4 & 2 & 90 \\
L8b / L96b & 8  & 4 & 4 & 3 & 454 \\
L8c / L96c & 12 & 6 & 6 & 2 & 189 \\
L8d / L96d & 12 & 6 & 6 & 3 & 1329\\
L96e & 16 & 8 & 8 & 3 & 2924\\
\hline\hline
\end{tabular}
\caption{List of runs. Each label encodes the dynamical system, either the extended Lorenz-type model of thermal convection (L8) or the Lorenz 96 atmospheric circulation model (L96). Configurations (a--e) of BSBS-RCM setup differ in the number of interferometer modes $N_{\rm mod}$, the number of photon sources $N_{\rm src}$, the number of vacuum entries $N_{\rm vac}=N_{\rm mod}-N_{\rm src}$, maximum photon number moment order $O_m$, and the number of observables $N_{\rm obs}$.}
\label{tab:runs}
\end{table}

\subsection{Memory and output reduction layer}
After the reduction of the sampled photon number measurements to expectation values of moments of the particle number, the vector of moments (here $O_m=2$), 
\begin{align}
    \bm r'(t+\Delta t)=(&\langle \hat n_1\rangle,\langle \hat n_2\rangle,...,\langle \hat n_{\tilde N}\rangle,\nonumber\\
    &\langle \hat n_1\hat n_1\rangle, \langle \hat n_1\hat n_2\rangle,...,\langle\hat n_{\tilde N}\hat n_{\tilde N}\rangle)^T
\end{align}
can serve as the nonlinear activation vector $\bm r'(t+\Delta t)$ for the next time step.
\begin{figure*}
    \centering
    \includegraphics[scale=0.72]{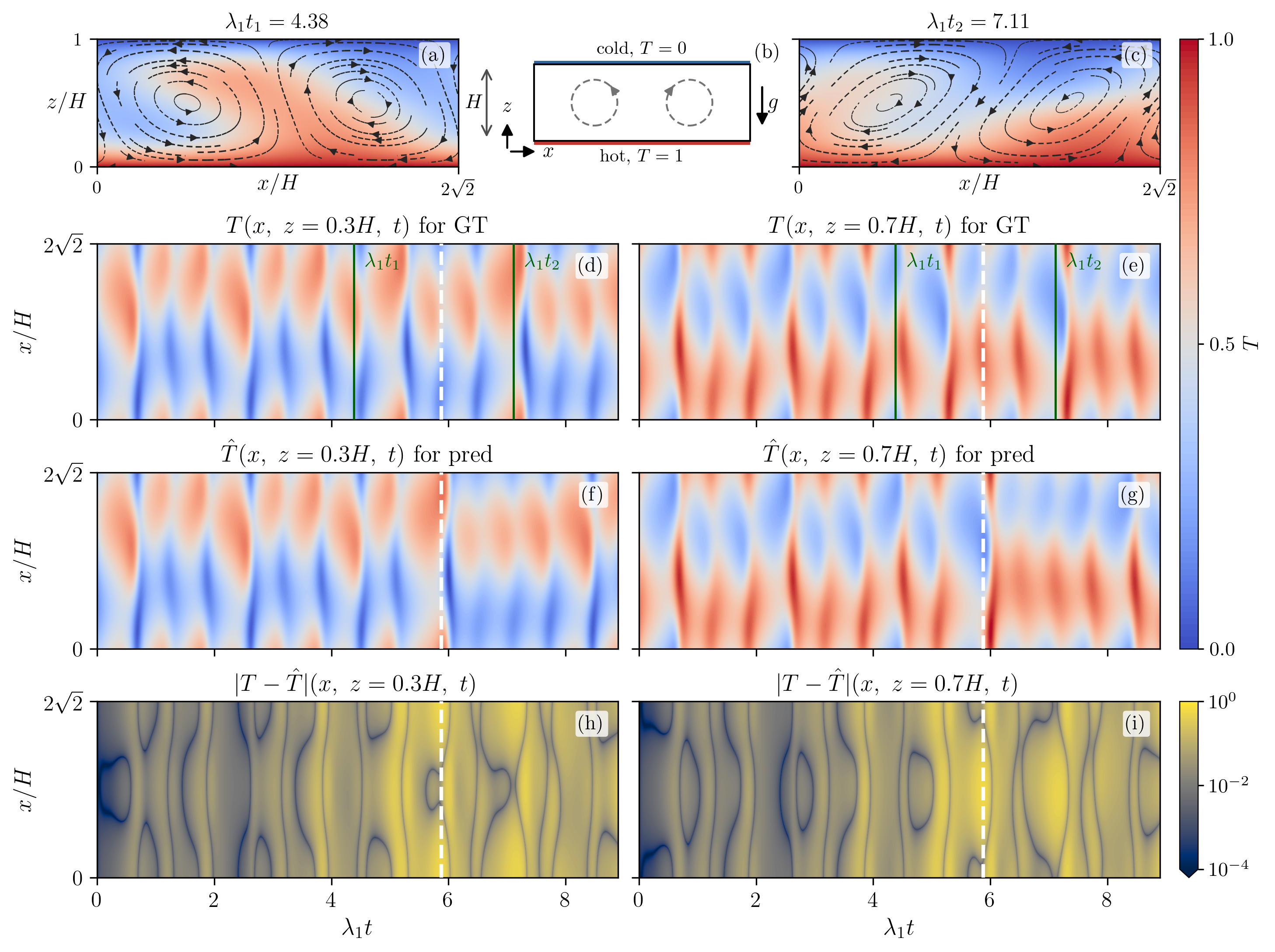}
    \caption{{\bf Prediction of the temperature in cuts through the convection layer.} (a) Convection flow snapshot at dimensionless time $\lambda_1 t =4.38$; $\lambda_1$ is the largest Lyapunov exponent of the dynamical system. (b) Principal sketch of the convection layer. (c) Further snapshot at $\lambda_1 t= 7.11$. We display the time evolution of the temperature field $T$ along two $x$-axes at $z/H = 0.3$ and 0.7, obtained from the prediction of the BSBS-RCM for the L8 model and compared to the ground truth (GT). Panels (d) and (e) show the GT of temperature $T$ in a space-time plot. Panels (f) and (g) show the corresponding prediction $\hat{T}$ from the multiphoton boson sampler (pred). Panels (h) and (i) display the difference between ground truth and prediction, $\vert T - \hat{T} \vert$. The color bar in the bottom row is given in logarithmic units. The example is taken for a particular quantum reservoir seed with $\varepsilon=0.02$, $N_{\rm mod}=8$, $N_{\rm src}=4$, and $O_m=2$ and gives a prediction horizon of $\lambda_1 T_p=5.9$, which is indicated by the thick white dashed line in the space-time plots. The solid vertical lines in (d) and (e) indicate the times of the snapshots in (a) and (c), respectively.}
    \label{fig:Fig3}
\end{figure*}
Our photonic reservoir computing model (BSBS-RCM) will be a hybrid quantum-classical algorithm since the memory is implemented classically outside the quantum photonic device in the next step of the data processing pipeline \cite{Pfeffer2022,Pfeffer2023}. The reservoir step at time $t+\Delta t$ is obtained by 
\begin{align}
	\label{eq:rcm_activation}
	{\bf r}(t+\Delta t)=(1-\varepsilon){\bm r}(t) + \varepsilon {\bm r}^{\prime}(t+\Delta t)\,.
\end{align}
The first term stands for a linear memory and the second term for the nonlinear activation. Note that the nonlinear activation follows from the encoding via the Hamiltonian \cite{Fujii2017,Pfeffer2022}. The leaking rate $\varepsilon$ blends between both contributions \cite{Inubushi2017}. For $\varepsilon=1$, one obtains again an extreme learning machine. In our case, this reduced information will re-enter the BSBS-RCM, similar to refs. \cite{Pfeffer2023,Kobayashi2024}. Finally, the reservoir state ${\bm r}(t+\Delta t)$ is fed into the optimized output layer $W_{\rm out}^*$ to give the state vector of the dynamical system at the next time step, 
\begin{equation}
{\hat{\bm X}}(t+\Delta t)=W_{\rm out}^* {\bm r}(t+\Delta t)\,.
\end{equation}
See again Fig.~\ref{fig:Fig1} for an illustration of the whole data processing pipeline. Rather than ridge regression, we use the pseudoinverse for most cases, since the multiphoton boson sampler output is still small in comparison to the number of training steps $N_{\rm train}/N_{\rm out} \gg 1$, see Appendix~\ref{app:PRCM_training} for further details.

\begin{table*} 
\begin{tabular}{lllll} 
\hline\hline
	 & $\qquad$& C-RCM & $\qquad$ & BSBS-RCM\\
\hline
Input vector size && $N_{\rm dof}$ & & $N_{\rm dof}$\\
Reservoir dimension $\dim_R$  && $10^2 N_{\rm dof}$ & & $(n_{\rm max}+1)^{2N_{\rm src}+N_{\rm vac}}$\\
Number of ``neurons" && $10^2N_{\rm dof}$ & & $2N_{\rm dof}$\\
Input operations && $N_{\rm dof}\times 10^2 N_{\rm dof}$ & & $N_{\rm dof}$\\
Reservoir operations && $10^2 N_{\rm dof}\times 10^2 N_{\rm dof}$ & & $N_{\rm dof}$\\
Output operations && $N_{\rm dof}\times 10^2 N_{\rm dof}$ & & $N_s$\\
Gate complexity $C_G$ && --- & & $ {\cal O}(N_{\rm dof}^2)$\\
Number of shots $N_s$ && 1 & & $1/\epsilon^2$\\
Query complexity $C_Q = C_G \times N_s$ && ${\cal O}(10^4 N^2_{\rm dof})$ & & ${\cal O}(N^2_{\rm dof}N_s)$\\
\hline\hline
\end{tabular}
\caption{List of the estimates for the (model) operations for one integration step of the classical dynamical system. We present complexity estimates for classical reservoir computing model (C-RCM) and the BSBS-based photonic reservoir computing model (BSBS-RCM). Here $N_{\rm src}$ is the number of photon sources, $N_{\rm vac}$ the number of vacuum modes. We denote by $N_s$ the number of shots or repetitions. The desired error threshold $\epsilon\sim 1/\sqrt{N_s}$.}
\label{tab:complexity_estimates} 
\end{table*}

\subsection{Estimate of number of operations}
Before presenting the prediction performance, it is pertinent to estimate the resources required in classical and photonic RCMs. The classical RCM lives typically in a space inflated by two orders of magnitude; thus we set $\dim_R\sim 10^2 N_{\rm dof}$. The size of the photonic reservoir is estimated as follows: when using $N_{\rm src}$ heralded Fock and $N_{\rm vac}$ additional vacuum states of photon number up to $n_{\rm max}$ then the dimensionality becomes $\dim_R\sim (n_{\rm max}+1)^{2N_{\rm src}+N_{\rm vac}}$. Single-photon sources ($n_{\rm max}=1$), as mostly used, would lead to a qubit-like exponential growth of the reservoir dimension, $2^{2N_{\rm src}+N_{\rm vac}}$. Multiphoton systems can grow significantly faster in their dimensionality. 

The loading into a classical reservoir requires ${\cal O}(10^2 N_{\rm dof}^2)$ operations. The advantage of BSBS-RCM is the potential dimensionality that can be encoded. With a $\sqrt{N_{\rm dof}}\times \sqrt{N_{\rm dof}}$ unitary $U_{\rm in}$, we can encode $N_{\rm dof}$ classical values. Thus we need ${\cal O}(N_{\rm dof})$ operations. The number of shots $N_s$ in case of BSBS-RCM remains somewhat uncertain. However, the repetition rates of experimental BSBS-RCMs in the 10 MHz regime will cause an advantage in the photonic case, even though it is not a universal quantum computer.

On the computational complexity, QML is a paradigm for data-based problem solving and approximation; it is not a computational problem in itself and therefore does not live in a specific complexity class. We do not make a statement of complexity theoretic advantage, rather a practical advantage on performing/encoding the data-processing algorithm for RCM, using a programmable quantum sampling machine -- a concept often referred to as \textit{quantum utility}~\cite{10234273}. Nevertheless, since BSBS-RCM is based on boson sampling, we know it can be quantum complex \cite{Aaronson2011}. 

Table~\ref{tab:complexity_estimates} summarizes our estimates. We see that both approaches scale with $N_{\rm dof}^2$ for the query complexity.

\section{Results}
\subsection{Prediction of thermal convection flow}
We present the results for two prediction tasks of complex, chaotic fluid flows.  The first task is for the two-dimensional thermal convection flow, which is described as a dynamical system in the form of a truncated Galerkin model \cite{Holmes2012}. The number of degrees of freedom in this model is $N_{\rm dof}=8$ \cite{Gluhovsky2002,Pfeffer2022}; it is thus an extended Lorenz-type model based on the classical work \cite{Lorenz1963} and will be termed L8 model in the following. The second prediction task is for a minimal model of global chaotic atmospheric circulation, the Lorenz 96 model \cite{Lorenz1995,Chattopadhyay2020}. This model contains all ingredients of the Navier-Stokes equations which typically describe turbulent fluid motion and thus geophysical turbulence: nonlinearity, external forcing, dissipation, spatial coupling, and chaos. The number of degrees of freedom is now $N_{\rm dof}=16$; it will be termed L96 model.

In thermal convection flows, buoyancy forces drive the complex fluid motion. The present system consists of a two-dimensional fluid layer, which is heated uniformly from below with a dimensionless rescaled temperature $T=1$ and cooled from above with a dimensionless $T=0$, thus giving an imposed temperature difference $\Delta T>0$. Once this thermal driving exceeds a threshold, which is expressed by the critical Rayleigh number parameter, fluid motion in the form of a pair of counter-rotating circulation rolls is initiated by a linear instability. Hot lighter fluid ascends, while cold heavier fluid descends, as shown in Fig.~\ref{fig:Fig3}(a,c). For sufficiently high Rayleigh number, the fluid motion becomes time-dependent and finally turbulent. The L8 model is given in compact form by
\begin{equation}
\frac{dX_i(\tau)}{d\tau}=\sum_{j,k=1}^{N_{\rm dof}} A_{ijk} X_j(\tau) X_k(\tau) + \sum_{j=1}^{N_{\rm dof}} B_{ij} X_j(\tau)\,. 
\label{eq:ODE-L8}
\end{equation}
This system of coupled ordinary differential equations is obtained by a Fourier mode expansion of velocity and temperature fields and subsequent truncation of the nonlinear partial differential Boussinesq equations that model a thermal convection flow. Here, $A_{ijk}$ and $B_{ij}$ are constant coefficients. It contains the quadratic nonlinearity that is characteristic for complex fluid flow problems. The ground truth is obtained by a numerical integration of \eqref{eq:ODE-L8} with a fourth-order Runge-Kutta scheme. The principal sketch of the layer geometry is shown in panel (b) of Fig.~\ref{fig:Fig3}. The full set of equations and its derivation are given in Appendix~\ref{app:derivation_convection_flow}.  

Figure~\ref{fig:Fig3}(d--i) displays the predictive capability of the BSBS-RCM for one example run. We show the space-time evolution of temperature field $T(x,t)$ at two fixed vertical locations: $z = 0.3$ in the bottom half and $z = 0.7$ in the top half as a space-time contour plot. Both predictions are obtained from a typical BSBS-RCM realization L8a (see again Table~\ref{tab:runs}). The predicted temperature fields $\hat{T}(x, t)$ at $z=0.3$ and 0.7 closely reproduce the ground truth $T(x, t)$ at first, very well capturing the spatial structure and temporal quasi-periodicity of the convective plumes. This includes the varying tilting of the thermal plumes---lumps of hot and cold fluid that detach from the bottom and top wall, respectively---in connection with a local shear flow motion of the circulation rolls, visible in panels (a) and (c) of Fig.~\ref{fig:Fig3}.

As a first performance metric, we take the normalized mean square error (NMSE), given by
\begin{align}
	\label{eq:mse}
	{\rm NMSE}(\tau)=\dfrac{\sum_{t'=1}^{\tau} \big|{\bm X}(t')-\hat{\bm X}(t')\big|^2}{\sum_{t'=1}^{\tau} {\bm X}^2(t')}\,.
\end{align}
Recall that we denote the BSBS-RCM prediction by $\hat{\bm X}$ and the ground truth by ${\bm X}$. In the closed-loop scenario, the predictive capability of the reservoir is characterised by the {\em valid prediction time} or {\em prediction horizon} $T_{\rm p}$ rather than by an NMSE \cite{Rathor2025}. The prediction horizon therefore defines a second metric and measures how accurately the evolution of characteristic structures of the fluid flow can be predicted. To evaluate this prediction horizon, one uses the time-dependent ${\rm NMSE}(t)$ from \eqref{eq:mse} and defines
\begin{equation}
    \label{eq:PH}
	\lambda_1 T_{\rm p} = \max\{\lambda_1 \tau \,|\, {\rm NMSE}(\lambda_1 t) < 10\%\,\forall t <\tau\}\,.
\end{equation}
Times are always measured in units of the maximal Lyapunov exponent $\lambda_1$ of the dynamical system, which carries the physical dimension of an inverse time and is given for the present set of system parameters, that is a Prandtl number $\sigma=10$, a relative Rayleigh number $r=28$, and a rescaled aspect ratio $b=8/3$ by $\lambda_1=0.89$ \cite{Rathor2026}. The largest Lyapunov exponent quantifies the chaoticity of the dynamics. The larger the number of degrees of freedom $N_{\rm dof}$, the larger typically $\lambda_1$ and thus sensitivity of the dynamics with respect to small differences in initial conditions. The parameters $(\sigma,r,b)$ are held constant in the present case study; see again Appendix~\ref{app:derivation_convection_flow}.

Figure~\ref{fig:Fig3} displays the time series for approximately 9 time units. It is clearly visible that BSBS-RCM captures the system trajectory for a time lag of 3 Lyapunov times with a normalized MSE $\lesssim 10^{-2}$. Even when the NMSE is further increased at later times, the overall qualitative dynamics is kept intact. The frequency and amplitude in the time series remain comparable, see again the space-time contour plots of Fig.~\ref{fig:Fig3}(d--g). The latter is essential for higher-dimensional turbulent systems with a stronger chaoticity. A step into this direction is done with the second use case of this study.

\begin{figure*}[t!]
\centering
\includegraphics[width = 0.95\textwidth]{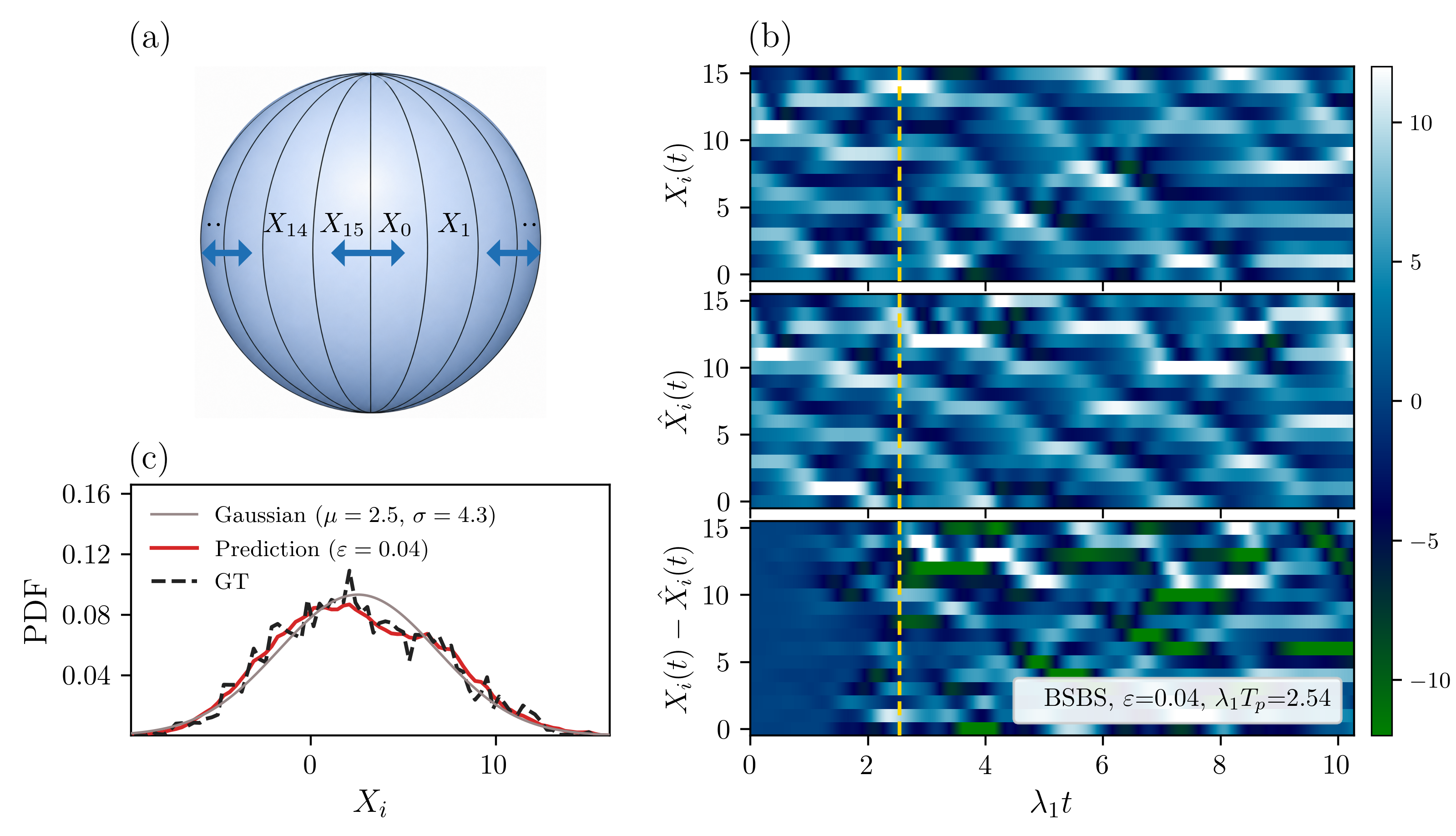}
\caption{{\bf Prediction of global circulation patterns with photonic reservoir computing model.} (a) Sketch of decomposition of the global atmosphere into sectors; each sector is represented by one variable $X_j$. Nonlinear coupling to the neighboring sectors generates traveling wave patterns. (b) One example case of the spatio-temporal evolution of the field. The top panel shows the ground truth, the middle panel shows the corresponding prediction by BSBS-RCM, and the bottom row displays the difference between the GT and the prediction. The color bar to the right holds for all three panels and gives the amplitude. The example is taken for a particular quantum reservoir seed at $\varepsilon=0.04$, $N_{\rm mod}=12$, $N_{\rm src}=6$, and $O_m=3$ and gives a prediction horizon of $\lambda_1 T_p=2.54$. (c) Reproduction of the statistical properties of L96. We show the probability density function (PDF) of $X_j$ for ground truth and prediction and compare the data with a Gaussian PDF.}
\label{fig:Fig4}
\end{figure*}
\begin{figure}[t!]
\centering
\includegraphics[scale = 0.9]{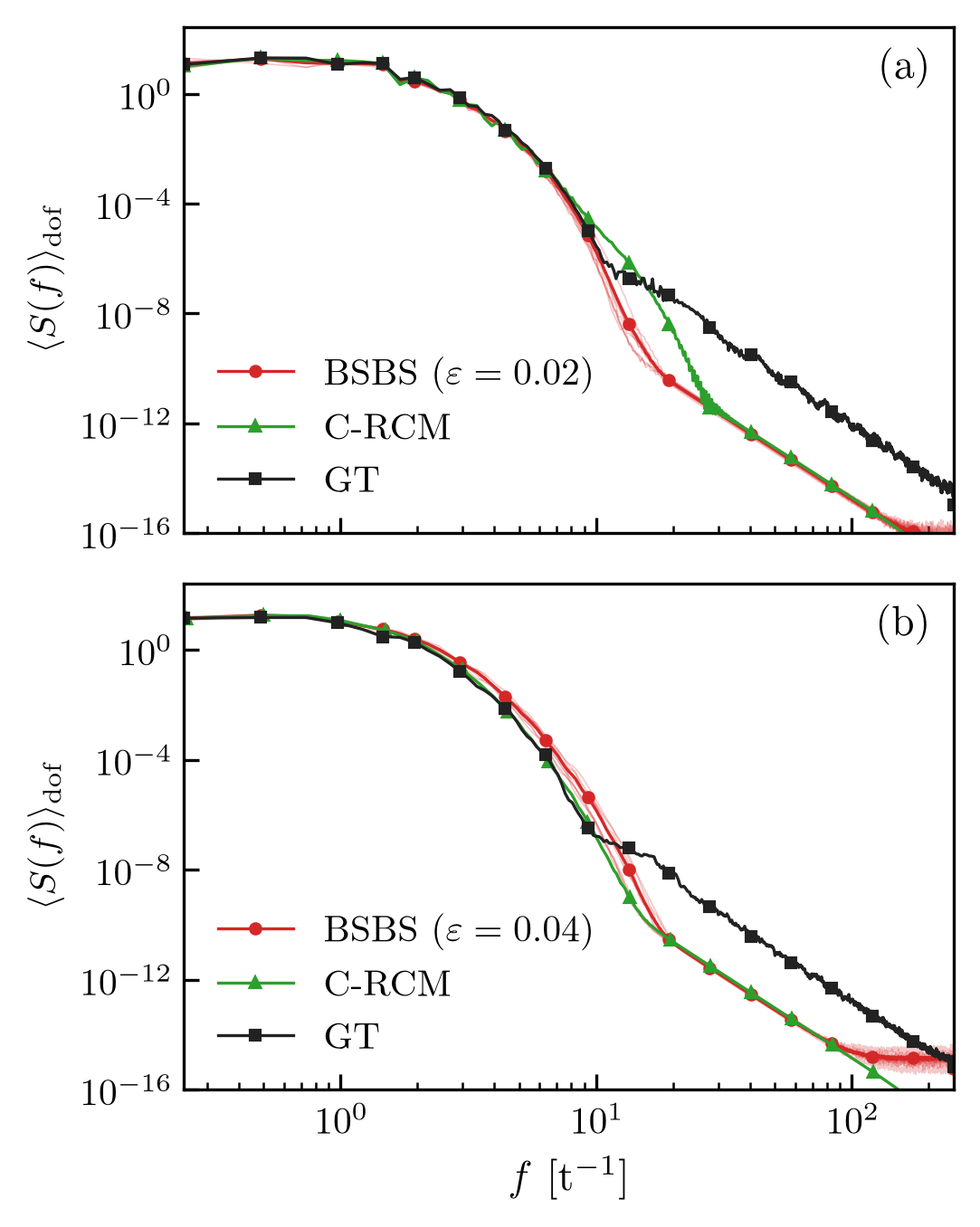}
\caption{{\bf Power spectra of ground truth and prediction.} (a) L8 model. (b) L96 model. The low-frequency range fits almost perfectly. Characteristically, both models, classical reservoir computing model (C-RCM) and bright scattershot boson sampling (BSBS), the photonic RCM, fail to predict high frequency signal components. Both cases are compared to the ground truth (GT).}
\label{fig:Fig5}
\end{figure}

\subsection{Prediction of atmospheric circulation}
The second prediction task is for the so-called Lorenz 96 model---a minimal model for atmospheric circulation processes on Earth. In L96, the variables $X_i(t)$ are arranged on a one-dimensional ring to represent $N_{\rm dof}$ sectors of the thin spherical-shell atmosphere layer as illustrated in Fig.~\ref{fig:Fig4}(a). The L96 model is given by the following system of nonlinear ordinary differential equations,
\begin{equation}
\frac{dX_i}{dt} = \left(X_{i+1}-X_{i-2}\right)X_{i-1} -X_i+F\,,
\end{equation}
with $F$ being a constant forcing term, $F=10$. Index $i=0,...,N_{\rm dof}-1$. The maximum Lyapunov exponent is now given by $\lambda_1=2.055$. The first term on the right hand side stands for the nonlinear advection, the driver of complex motion. The second linear term represents dissipation and friction effects. We assume periodic boundary conditions, $X_{i+N_{\rm dof}}=X_i$. Figure~\ref{fig:Fig4}(b) shows a typical example of L96 dynamics in a space-time contour plot. The upper panel presents the ground truth dynamics, the middle panel the BSBS-RCM prediction. The bottom contour plot of panel (b) displays the difference between both. The prediction horizon $\lambda_1 T_p$ of the ensemble study (which will be discussed in detail in the next section) is inserted as a dashed yellow line in all panels. The BSBS-RCM can predict more than two Lyapunov times on average before the relative error exceeds the 10\% threshold. 

Even beyond the prediction horizon, the statistical properties of the system are accurately reproduced by the BSBS-RCM. This is illustrated in Fig.~\ref{fig:Fig4}(c), which shows the probability density functions (PDF) of the variables $X_i$. Although an individual predicted trajectory inevitably diverges from the ground truth after a few Lyapunov times, the underlying statistical properties remain well reproduced. The PDF is taken over the full test time interval. 

This observation is further supported by Fig.~\ref{fig:Fig5}, where we compare the power spectra of the ground truth and the RCM predictions for both use cases in doubly logarithmic plots. At the highest frequencies, corresponding to the shortest time scales, neither the classical nor the photonic RCM fully captures the dynamics quantitatively; spectra decay faster with increasing $f$. However, qualitatively the decay is reproduced well.  In contrast, the lower-frequency and more slowly varying components are reproduced very well. Capturing these large-scale statistical properties and long-time behavior is an important prerequisite for using the model as a surrogate for complex fluid flow dynamics.

\begin{figure}[t!]
\centering
\includegraphics[scale = 0.9]{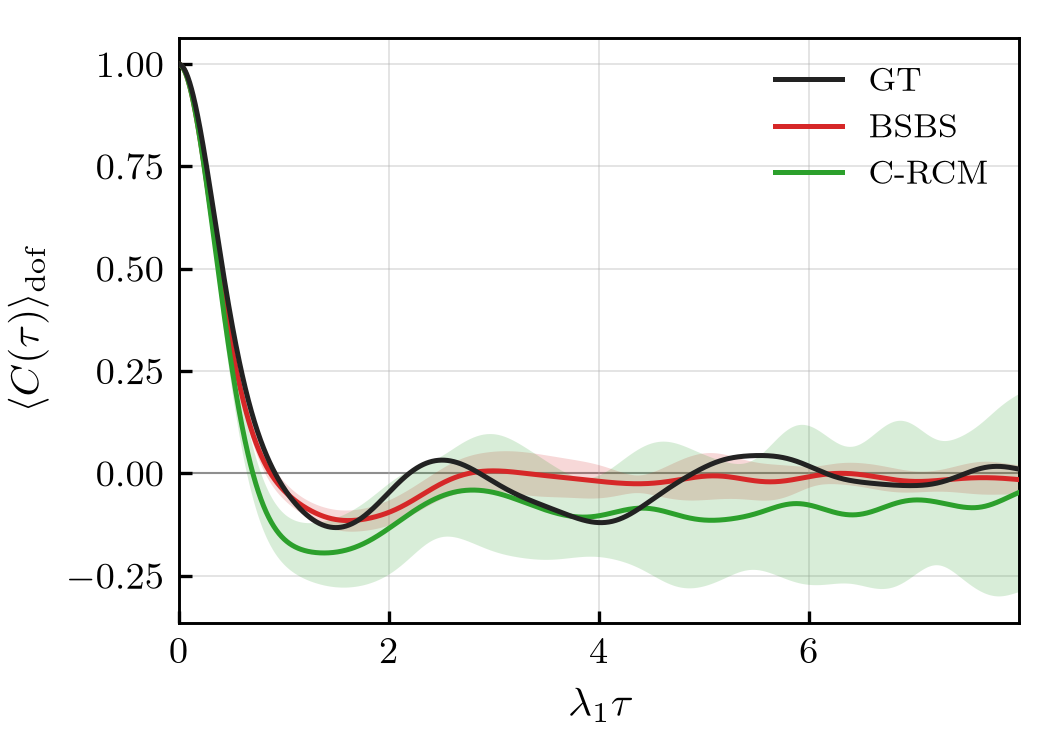}
\caption{{\bf Time-correlations of L96 model.} The time correlation function $\langle C(\tau)\rangle_{\rm dof}$, which is an average over the individual correlation function of the degrees of freedom, is plotted versus delay time $\tau $.  Shaded bands for BSBS and C-RCM denote variance over reservoir seeds. The long-time integral correlation time is $\tau_{\rm int}=0.22$ for the ground truth (GT, black solid line); $\tau_{\rm int}=0.21\pm 0.07$ for BSBS (red solid line), and $|\tau_{\rm int}|= 0.33\pm 0.18$ for C-RCM (green solid line).}
\label{fig:time_corr_l96_dofmean}
\end{figure}

Figure~\ref{fig:time_corr_l96_dofmean} shows the time-correlation function $\langle C(\tau)\rangle_{\rm dof}$ averaged over all 16 degrees of freedom of L96, comparing the BSBS and classical RCM (which we term C-RCM) predictions with the ground truth. All three curves decay from unity over a similar timescale and the ground truth exhibits a weak recurrence, visible as a small secondary peak near $\lambda_1\tau\approx2.5$ that recedes and reappears at longer lags.  We summarize this behavior with the long-time integral correlation time $\tau_{\rm int}=\int_0^{\tau_{\max}} \langle C(\tau)\rangle_{\rm dof}\,d\tau$, with $\lambda_1\tau_{\max}=8$. BSBS reproduces the integral time closely with $\tau_{\rm int}=0.21\pm0.07$ compared to the ground truth value of $\tau_{\rm int}=0.22$. The C-RCM overshoots in magnitude after the initial decay and does not recover the oscillations, yielding an integral time of $\tau_{\rm int}=-0.33\pm0.18$. Unlike the power spectral density, which carries the same information in the frequency domain, or the single-time probability distribution, which carries no temporal information at all, $\langle C(\tau)\rangle_{\rm dof}$ probes the dynamical memory of the surrogate model: how strongly a variable's current state constrains its own future, and whether that memory is structured (e.g. recurrent) rather than purely decaying. The close agreement of  BSBS  with the ground truth indicates that it reproduces not only the correct aggregate statistics, but also the temporal structure of the attractor's internal dynamics more faithfully than C-RCM.

\subsection{Prediction capability for different reservoir seeds and photon loss rates}
We showed the prediction capability of BSBS-RCM for two use-cases. In the following, we still have to investigate the robustness of these results with respect to hyperparameters of the reservoir computing models. In the photonic system, these are the size of the reservoir, which is mainly determined by the number of photon sources, $N_{\rm src}$, the number of interferometer modes, $N_{\rm mod}$, the maximum order of photon number moments, $O_m$, and the leaking rate $\varepsilon$. Table~\ref{tab:runs} summarizes the parametric studies that were conducted. We used $N_{\rm src}=N_{\rm mod}/2$ for all photonic reservoir model runs. 

For each set of parameters of L8 and L96, we run an ensemble of 70 different random seeds of the unitary $U_{\rm rand}[\bm{\phi}]$ to calculate means and standard deviations of the prediction horizon. In addition, for each parameter set and each random seed, we ran the model for 50 different leaking rates $0<\varepsilon\le 1$ in increments of $\Delta\varepsilon=0.02$. We verified that the prediction horizon $\lambda_1 T_p=0$ for $\varepsilon=0$. This sums up to a total of 3500 runs for each BSBS-RCM setup. Each of these runs proceeds through a training phase of $3\times 10^4$ time steps $\Delta t=0.002$; the testing phase consists of $1\times 10^4$ steps. All runs were executed by simulations on a CPU cluster with 192 CPU cores, applying the Python-based photonic simulation software package Strawberry Fields~\cite{strawberryfields,Bromley_2020} for photonic circuits in combination with The Walrus library~\cite{Gupt2019}. The photon-number moments were calculated numerically exactly. 

We also compare the results of BSBS-RCM with a classical RCM, which went through a comprehensive hyperparameter tuning by a systematic grid search \cite{Rathor2025,Rathor2026}. Details of this optimization are contained in Appendix~\ref{app:CRCM}. 

\begin{figure*}
\includegraphics[width = 0.95\textwidth]{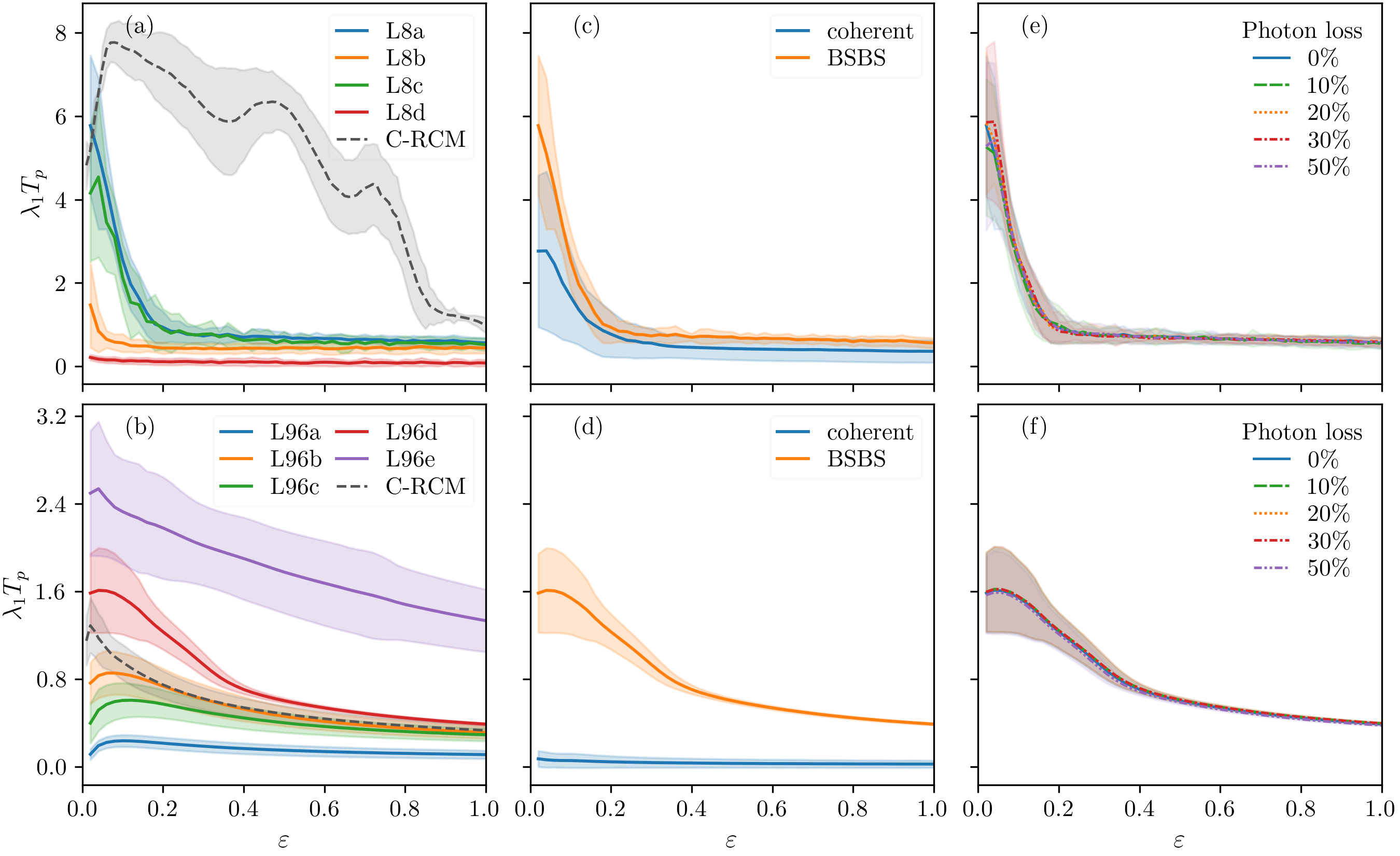}
\caption{{\bf Ensemble analysis of photonic reservoir computing model.} (a) Prediction horizon $\lambda_1 T_p$ for different leaking rates $\varepsilon$ and comparison with a classical RCM of comparable size for the L8 model. We used 50 different leaking rates $0 < \varepsilon \le 1$, starting with $\varepsilon=0.02$ in equidistant steps. Each data point is obtained from an ensemble of 70 different reservoir seeds. The solid lines connect the means, the margins indicate the variance of the statistics. The different L8 model runs are also listed in Table~\ref{tab:runs}. The black dashed lines in both figures stand for the classical RCM that was optimized by a comprehensive hyperparameter grid search. (b) Same as (a) for the L96 model runs. (c) Comparison of  bright scattershot boson sampling BSBS with coherent light run at exactly the same parameter set for L8. (d) Same as (c) for L96. (e) Prediction performance for different photon loss rates that are indicated in the legend. Analysis is for L8. (f) Same as (e) for L96.}
\label{fig:Fig6}
\end{figure*}

Figure~\ref{fig:Fig6} summarizes the results. Panel (a) shows the prediction horizons for L8. The solid lines represent the ensemble means, while the surrounding shaded regions indicate the corresponding standard deviation. Recall that the prediction horizon is defined as the time at which the NMSE first exceeds 10\%. The best performance is obtained for small leaking rates, whereas the prediction horizon decreases with increasing $\varepsilon$. In particular, run L8a achieves the best performance for $N_{\rm mod}=8$, $N_{\rm src}=4$, and $O_m=2$ at $\varepsilon=0.02$. This corresponds to the regime in which the contribution of the linear photonic network to the reservoir update is minimal. For the classical RCM, the prediction horizon is approximately 25\% larger, with the optimum slightly shifted to $\varepsilon=0.06$. Interestingly, Pfeffer et al. \cite{Pfeffer2022} observed a similar dependence on the leaking rate for a qubit-based RCM, with the best prediction performance likewise obtained at small values of $\varepsilon$. It underlines the importance of the linear memory.

Furthermore, we observe that an increase in the number of photon sources, modes, and of the order of the photon number moment does not improve the prediction performance of the L8 case. The performance actually drops significantly for larger reservoir sizes. In the case of L8, the photonic RCM does not outperform a highly optimized classical RCM, the latter of which obtains a mean prediction horizon of almost 8 Lyapunov times.
However, BSBS-RCM still captures the dynamics for up to 6 Lyapunov times.

Panel (b) of Fig.~\ref{fig:Fig6} summarizes the results for L96; now the classical case is clearly outperformed by the photonic one. The mean prediction horizon drops to more than 2.5 Lyapunov times for this higher-dimensional use case, it is by about a factor of 2 better than the classical one. The best ensemble members reach more than 3 Lyapunov times. The higher dimensionality of the prediction task ($N_{\rm dof}=16$) requires a higher-dimensional photonic reservoir and a higher maximum moment order. The best performance is now in line with an increased leaking rate of $\varepsilon=0.04$, suggesting a slightly stronger contribution of the boson sampler to the update step. Our results confirm a well-known circumstance of reservoir computing models: the choice of the system parameters is task-specific. This also shows how we can make use of the genuine high dimensionality of the photonic system for complex learning tasks.

Next, we investigated the robustness of the BSBS-RCM performance for different rates of photon loss with our digital twin. They are present in each experiment and thus an intrinsic property of real photonic reservoir computing devices. We have evaluated different loss rates at the inputs and outputs of up to 50\% (see also Appendix~\ref{app:PRCM}). Panels (e) and (f) of Fig.~\ref{fig:Fig6} summarize our results for model runs L8a and L96d, respectively. While there is a small impact for model L8a, we do not observe any difference for L96d. These results indicate a high degree of resilience to photon loss. 

In summary, we can state that our results demonstrate clearly that an application case that is sufficiently high-dimensional outperforms the corresponding classical algorithm even with included errors -- a promising route to photonic quantum advantage.

\subsection{Comparison to runs with coherent light}\label{sec:coherent}
\begin{figure}
\includegraphics[width = 0.48\textwidth]{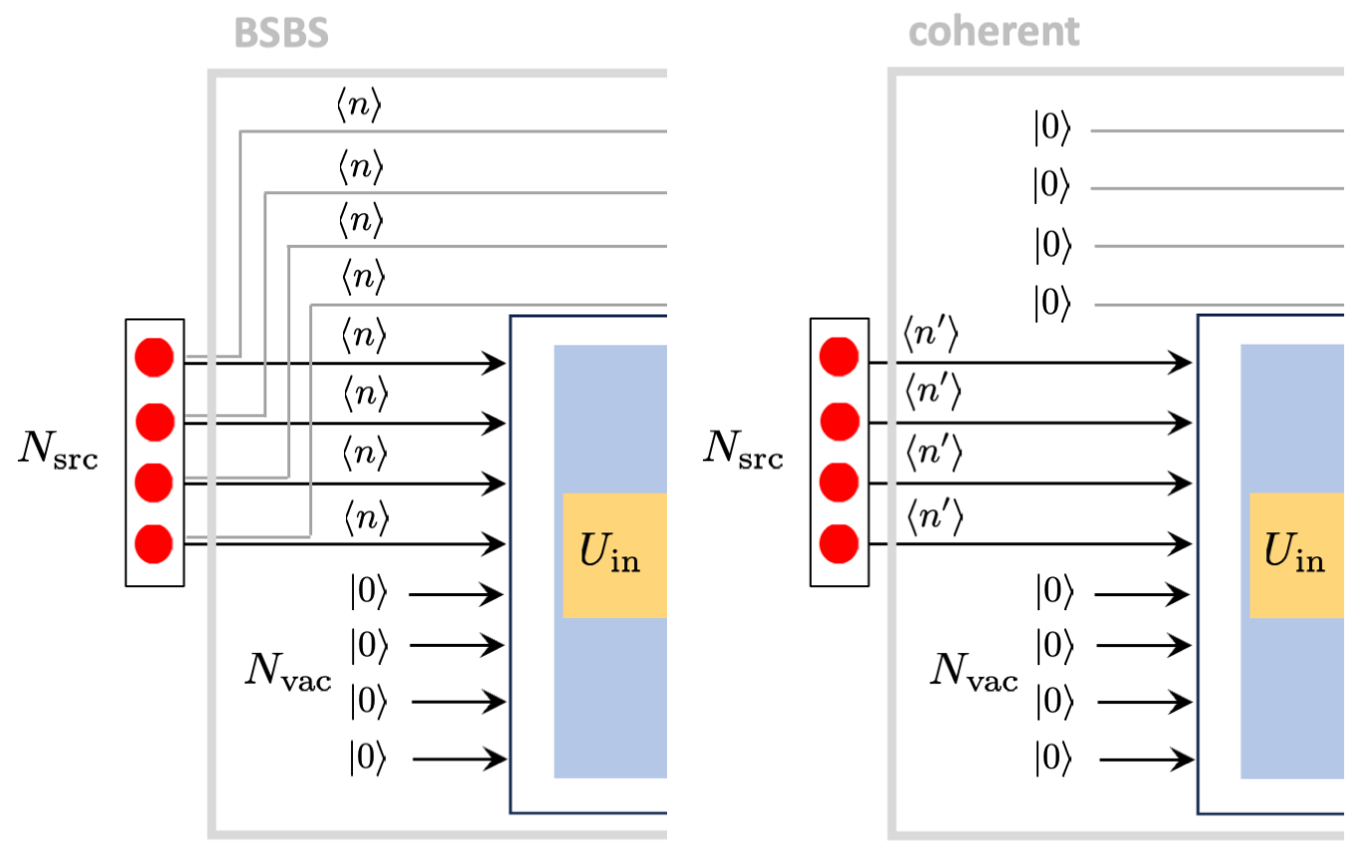}
\caption{{\bf Comparison with coherent light.} Sketch of the photon input for bright scattershot boson sampling (BSBS) and the experiment with coherent light for consistent comparison.}
\label{fig:Fig7}
\end{figure}

To further validate the conclusions drawn from the previous subsection, we simulate the system using coherent light, which serves as a heuristic benchmark for a classical analog of the system. To this end, we take the best-performing BSBS-RCM configurations and simulate both use cases with coherent-light input. Figure~\ref{fig:Fig7} illustrates the corresponding BSBS-RCM configurations at input to obtain a consistent comparison between both cases. The average photon number in the BSBS case is $\langle n\rangle=0.73$ for the modes that enter the interferometer and heralding arms. In the case of coherent light, the mean photon number of the modes that enter the interferometer has to be doubled, $\langle n'\rangle=2\langle n\rangle$ to provide a fair comparison -- i.e. total optical power remains constant, although we do note this also means that double the power enters the interferometer. As in the photonic quantum cases, we use the same number of random seeds and the same set of leaking-rate values. 

The corresponding results are shown in panels (c) and (d) of Fig.~\ref{fig:Fig6}. They clearly demonstrate that the reservoir computing model performs better with quantum light than with classical coherent light. This is in particular the case for the higher-dimensional Lorenz 96 model for the atmospheric circulation. 

\section{Discussion}
The results shown above are based on an exact numerical simulation of the proposed BSBS setup. The simulation uses Gaussian states, which allow an efficient realization of the state generation, action of the interferometer, and losses. At the detection stage, photon-number moments are obtained exactly from expectation values of the Gaussian state. This step includes the intrinsic computational complexity, as it requires the exponentially scaling calculation of Hafnians with dimensions linear in the order of the photon number moments $O_m$.

Compared to the experimental setup, where both training and subsequent prediction would be executed \textit{in situ}, i.e., directly on the experimental setup, a number of effects are not yet considered in our digital twin simulations: The transition from directly calculated expectation values to sampled photon number moments contributes statistical noise, i.e., shot noise, which could affect the training quality if too few samples are used. The BSBS approach has been designed, such that processing of the samples is straightforward. The reduction of the samples to expectation values can be performed directly, for example, by an FPGA close to the detectors without storing the samples explicitly.
Another effect not considered in the simulations above is the saturation of detectors, where, for example, typical superconducting nanowire single-photon detectors can only detect a limited number of photons per event. This saturation behavior does not impact the viability of the proposed setup because it does not require the photon number moments to be precise. Instead, the relation between dynamical inputs $\bm X$ and the sampled photon number moments $(\langle \hat n_1 \rangle_\approx, \langle \hat n_2 \rangle_\approx,...,\langle \hat n_1\hat n_1 \rangle_\approx, \langle \hat n_1\hat n_2 \rangle_\approx,...)$ must be sufficiently nonlinear.

In addition to neglecting shot noise and detector saturation, an ideal programmable interferometer was assumed. Real programmable interferometers only have a limited programming fidelity, i.e., the implemented unitary only approximately resembles the desired unitary. However, the proposed scheme does not rely on an exact implementation of the unitary, but the main requirement is that the programming be repeatable with sufficient accuracy and that close unitaries result in close interferometer characteristics, i.e., a continuous mapping. This guarantees that if a unitary occurs during prediction that is similar to a unitary used during training, it will result in a similar interferometer behavior and, thus, a faithful prediction. In general, we expect the proposed approach to be resilient against systematic and repeatable programming inaccuracies of the interferometer but sensitive to time-dependent drifts during training or prediction.

In summary, we have demonstrated that a multiphoton boson sampler, a hybrid quantum-classical quantum machine learning algorithm, is capable of predicting complex fluid flow dynamics. When scaling up the problem size --- which implies here to go from the thermal convection flow model to the atmospheric circulation model --- we obtained a significantly better performance in comparison to a classical reservoir computing model. We showed in addition that our digital twin of the Paderborn Quantum Sampler is robust with respect to photon loss rates up to 50\%. The algorithm on the basis of quantum light outperforms furthermore a comparable configuration with coherent light.  These results demonstrate an advantage in using the quantum states considered here rather than coherent states, consistent with the expectation that the former exhibit more complex photon-number correlations. The implementation of the present simulations as a real experiment and a further up-scaling of the problem size are currently underway and will be reported elsewhere.

\acknowledgments
The work of SKR and JS is supported by the European Union (ERC, MesoComp, 101052786). Views and opinions expressed are however those of the author only and do not necessarily reflect those of the European Union or the European Research Council. Neither the European Union nor the granting authority can be held responsible for them. The authors gratefully acknowledge the computing time made available to them on the high-performance computer Otus~\cite{ehtesabi2025otussupercomputer} at the NHR Center Paderborn Center for Parallel Computing (PC2). This center is jointly supported by the Federal Ministry of Research, Technology and Space and the state governments participating in the National High-Performance Computing (NHR) joint funding program (www.nhr-verein.de/en/our-partners). This work has received funding from the German Federal Ministry of Research, Technology and Space within the PhoQuant project (Grant No. 13N16103).

\section*{Appendixes}
The following appendices provide technical details on the digital twin algorithm for the BSBS-based reservoir computing model (BSBS-RCM) in Appendix~\ref{app:PRCM}, a derivation of the extended Lorenz-type model from the Navier-Stokes equations including definitions of parameters and scales in Appendix~\ref{app:derivation_convection_flow}, and details on the classical reservoir computing model (C-RCM), which was used for comparison, in Appendix~\ref{app:CRCM}.

\appendix
\section{Implementation of digital twin}\label{app:PRCM}
\subsection{Architectures}
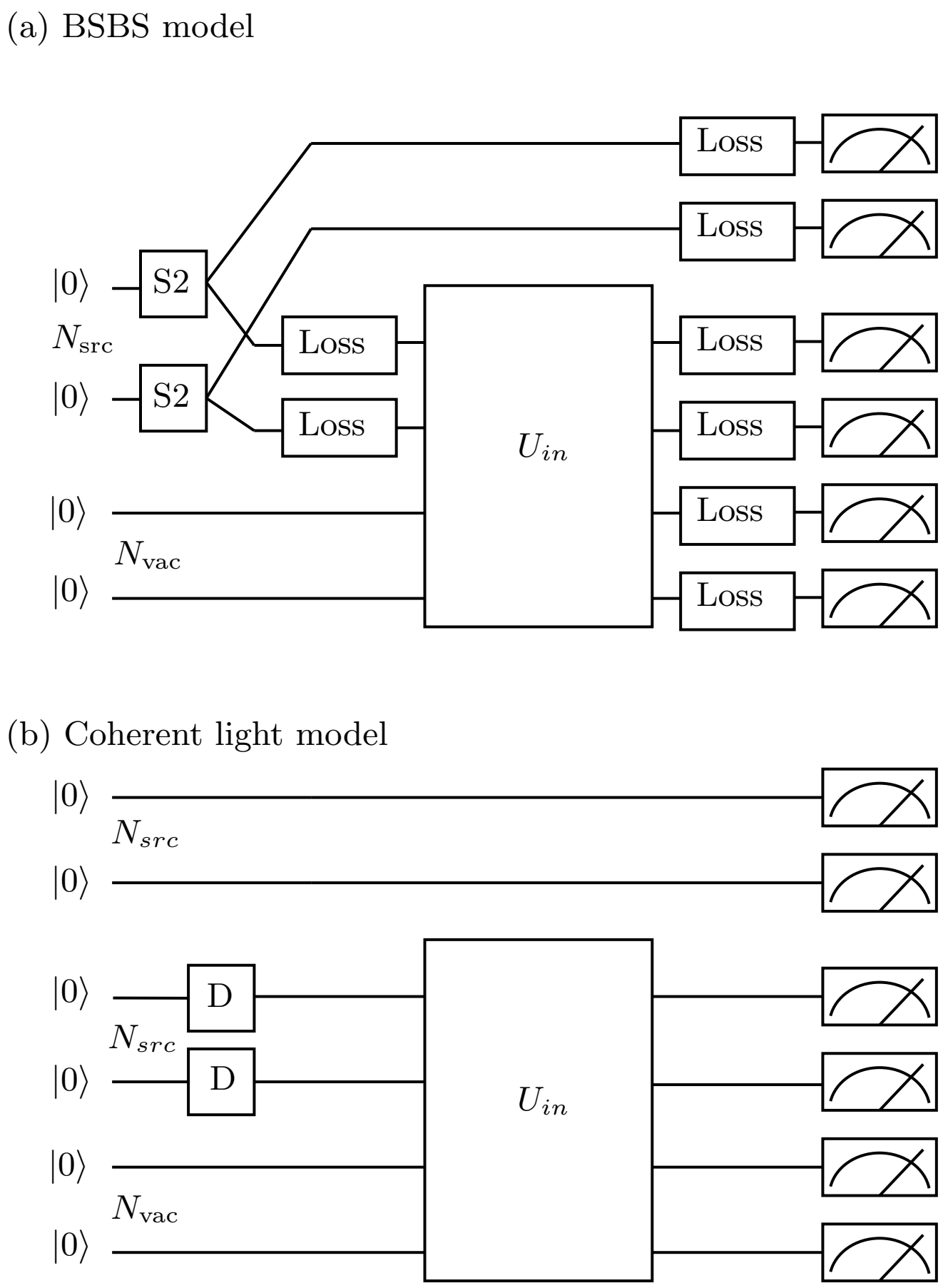
\begin{figure}

\tikzset{every picture/.style={line width=0.75pt}} 

\begin{tikzpicture}[x=0.75pt,y=0.75pt,yscale=-1,xscale=1]

\draw   (280,120) -- (360,120) -- (360,240) -- (280,240) -- cycle ;
\draw    (240,100) -- (370,100) ;
\draw   (420,60) -- (460,60) -- (460,80) -- (420,80) -- cycle ;
\draw    (440,80) -- (455.33,63.83) ;
\draw  [draw opacity=0] (422.74,76.8) .. controls (424.46,70.05) and (431.54,65) .. (440,65) .. controls (448.72,65) and (455.96,70.36) .. (457.41,77.42) -- (440,80) -- cycle ; \draw   (422.74,76.8) .. controls (424.46,70.05) and (431.54,65) .. (440,65) .. controls (448.72,65) and (455.96,70.36) .. (457.41,77.42) ;  
\draw   (370,61) -- (410,61) -- (410,81) -- (370,81) -- cycle ;
\draw   (180,148) -- (203.17,148) -- (203.17,171.17) -- (180,171.17) -- cycle ;
\draw   (230,131) -- (270,131) -- (270,151) -- (230,151) -- cycle ;
\draw   (180,107.83) -- (203.17,107.83) -- (203.17,131) -- (180,131) -- cycle ;
\draw    (203.17,118.83) -- (240,70) ;
\draw    (203.17,118.83) -- (220,141) ;
\draw    (220,141) -- (230,141) ;
\draw    (220,171) -- (230,171) ;
\draw    (220,171) -- (203.17,159.58) ;
\draw    (203.17,159.58) -- (240,100) ;
\draw    (170,160) -- (180,160) ;
\draw    (170,121) -- (180,121) ;
\draw    (410,69.67) -- (420,69.67) ;
\draw    (360,140) -- (370,140) ;
\draw    (360,171) -- (370,171) ;
\draw    (360,200) -- (370,200) ;
\draw    (360,230) -- (370,230) ;
\draw    (270,140) -- (280,140) ;
\draw    (270,170) -- (280,170) ;
\draw    (170,200) -- (280,200) ;
\draw    (170,230) -- (280,230) ;
\draw   (230,160) -- (270,160) -- (270,180) -- (230,180) -- cycle ;
\draw    (240,70) -- (370,70) ;
\draw   (420,90) -- (460,90) -- (460,110) -- (420,110) -- cycle ;
\draw    (440,110) -- (455.33,93.83) ;
\draw  [draw opacity=0] (422.74,106.8) .. controls (424.46,100.05) and (431.54,95) .. (440,95) .. controls (448.72,95) and (455.96,100.36) .. (457.41,107.42) -- (440,110) -- cycle ; \draw   (422.74,106.8) .. controls (424.46,100.05) and (431.54,95) .. (440,95) .. controls (448.72,95) and (455.96,100.36) .. (457.41,107.42) ;  
\draw   (370,91) -- (410,91) -- (410,111) -- (370,111) -- cycle ;
\draw    (410,99.67) -- (420,99.67) ;
\draw   (420,130) -- (460,130) -- (460,150) -- (420,150) -- cycle ;
\draw    (440,150) -- (455.33,133.83) ;
\draw  [draw opacity=0] (422.74,146.8) .. controls (424.46,140.05) and (431.54,135) .. (440,135) .. controls (448.72,135) and (455.96,140.36) .. (457.41,147.42) -- (440,150) -- cycle ; \draw   (422.74,146.8) .. controls (424.46,140.05) and (431.54,135) .. (440,135) .. controls (448.72,135) and (455.96,140.36) .. (457.41,147.42) ;  
\draw   (370,131) -- (410,131) -- (410,151) -- (370,151) -- cycle ;
\draw    (410,139.67) -- (420,139.67) ;
\draw   (420,160) -- (460,160) -- (460,180) -- (420,180) -- cycle ;
\draw    (440,180) -- (455.33,163.83) ;
\draw  [draw opacity=0] (422.74,176.8) .. controls (424.46,170.05) and (431.54,165) .. (440,165) .. controls (448.72,165) and (455.96,170.36) .. (457.41,177.42) -- (440,180) -- cycle ; \draw   (422.74,176.8) .. controls (424.46,170.05) and (431.54,165) .. (440,165) .. controls (448.72,165) and (455.96,170.36) .. (457.41,177.42) ;  
\draw   (370,161) -- (410,161) -- (410,181) -- (370,181) -- cycle ;
\draw    (410,169.67) -- (420,169.67) ;
\draw   (420,190) -- (460,190) -- (460,210) -- (420,210) -- cycle ;
\draw    (440,210) -- (455.33,193.83) ;
\draw  [draw opacity=0] (422.74,206.8) .. controls (424.46,200.05) and (431.54,195) .. (440,195) .. controls (448.72,195) and (455.96,200.36) .. (457.41,207.42) -- (440,210) -- cycle ; \draw   (422.74,206.8) .. controls (424.46,200.05) and (431.54,195) .. (440,195) .. controls (448.72,195) and (455.96,200.36) .. (457.41,207.42) ;  
\draw   (370,191) -- (410,191) -- (410,211) -- (370,211) -- cycle ;
\draw    (410,199.67) -- (420,199.67) ;
\draw   (420,220) -- (460,220) -- (460,240) -- (420,240) -- cycle ;
\draw    (440,240) -- (455.33,223.83) ;
\draw  [draw opacity=0] (422.74,236.8) .. controls (424.46,230.05) and (431.54,225) .. (440,225) .. controls (448.72,225) and (455.96,230.36) .. (457.41,237.42) -- (440,240) -- cycle ; \draw   (422.74,236.8) .. controls (424.46,230.05) and (431.54,225) .. (440,225) .. controls (448.72,225) and (455.96,230.36) .. (457.41,237.42) ;  
\draw   (370,221) -- (410,221) -- (410,241) -- (370,241) -- cycle ;
\draw    (410,229.67) -- (420,229.67) ;
\draw   (280,350) -- (360,350) -- (360,470) -- (280,470) -- cycle ;
\draw    (240,330) -- (420,330) ;
\draw   (420,290) -- (460,290) -- (460,310) -- (420,310) -- cycle ;
\draw    (440,310) -- (455.33,293.83) ;
\draw  [draw opacity=0] (422.74,306.8) .. controls (424.46,300.05) and (431.54,295) .. (440,295) .. controls (448.72,295) and (455.96,300.36) .. (457.41,307.42) -- (440,310) -- cycle ; \draw   (422.74,306.8) .. controls (424.46,300.05) and (431.54,295) .. (440,295) .. controls (448.72,295) and (455.96,300.36) .. (457.41,307.42) ;  
\draw    (170,300) -- (240,300) ;
\draw    (170,330) -- (240,330) ;
\draw    (360,370) -- (420,370) ;
\draw    (360,401) -- (420,401) ;
\draw    (360,430) -- (420,430) ;
\draw    (360,460) -- (420,460) ;
\draw    (220,370) -- (248.33,370) -- (280,370) ;
\draw    (220,400) -- (280,400) ;
\draw    (170,430) -- (280,430) ;
\draw    (170,460) -- (280,460) ;
\draw    (240,300) -- (420,300) ;
\draw   (420,320) -- (460,320) -- (460,340) -- (420,340) -- cycle ;
\draw    (440,340) -- (455.33,323.83) ;
\draw  [draw opacity=0] (422.74,336.8) .. controls (424.46,330.05) and (431.54,325) .. (440,325) .. controls (448.72,325) and (455.96,330.36) .. (457.41,337.42) -- (440,340) -- cycle ; \draw   (422.74,336.8) .. controls (424.46,330.05) and (431.54,325) .. (440,325) .. controls (448.72,325) and (455.96,330.36) .. (457.41,337.42) ;  
\draw   (420,360) -- (460,360) -- (460,380) -- (420,380) -- cycle ;
\draw    (440,380) -- (455.33,363.83) ;
\draw  [draw opacity=0] (422.74,376.8) .. controls (424.46,370.05) and (431.54,365) .. (440,365) .. controls (448.72,365) and (455.96,370.36) .. (457.41,377.42) -- (440,380) -- cycle ; \draw   (422.74,376.8) .. controls (424.46,370.05) and (431.54,365) .. (440,365) .. controls (448.72,365) and (455.96,370.36) .. (457.41,377.42) ;  
\draw   (420,390) -- (460,390) -- (460,410) -- (420,410) -- cycle ;
\draw    (440,410) -- (455.33,393.83) ;
\draw  [draw opacity=0] (422.74,406.8) .. controls (424.46,400.05) and (431.54,395) .. (440,395) .. controls (448.72,395) and (455.96,400.36) .. (457.41,407.42) -- (440,410) -- cycle ; \draw   (422.74,406.8) .. controls (424.46,400.05) and (431.54,395) .. (440,395) .. controls (448.72,395) and (455.96,400.36) .. (457.41,407.42) ;  
\draw   (420,420) -- (460,420) -- (460,440) -- (420,440) -- cycle ;
\draw    (440,440) -- (455.33,423.83) ;
\draw  [draw opacity=0] (422.74,436.8) .. controls (424.46,430.05) and (431.54,425) .. (440,425) .. controls (448.72,425) and (455.96,430.36) .. (457.41,437.42) -- (440,440) -- cycle ; \draw   (422.74,436.8) .. controls (424.46,430.05) and (431.54,425) .. (440,425) .. controls (448.72,425) and (455.96,430.36) .. (457.41,437.42) ;  
\draw   (420,450) -- (460,450) -- (460,470) -- (420,470) -- cycle ;
\draw    (440,470) -- (455.33,453.83) ;
\draw  [draw opacity=0] (422.74,466.8) .. controls (424.46,460.05) and (431.54,455) .. (440,455) .. controls (448.72,455) and (455.96,460.36) .. (457.41,467.42) -- (440,470) -- cycle ; \draw   (422.74,466.8) .. controls (424.46,460.05) and (431.54,455) .. (440,455) .. controls (448.72,455) and (455.96,460.36) .. (457.41,467.42) ;  
\draw   (196.83,359) -- (220,359) -- (220,382.17) -- (196.83,382.17) -- cycle ;
\draw   (196.83,388.42) -- (220,388.42) -- (220,411.58) -- (196.83,411.58) -- cycle ;
\draw    (170.33,370.5) -- (196.83,370.58) ;
\draw    (170,400) -- (196.83,400) ;

\draw (311,170.4) node [anchor=north west][inner sep=0.75pt]    {$U_{in}$};
\draw (374,63) node [anchor=north west][inner sep=0.75pt]   [align=left] {Loss};
\draw (182.83,153) node [anchor=north west][inner sep=0.75pt]   [align=left] {S2};
\draw (234,134) node [anchor=north west][inner sep=0.75pt]   [align=left] {Loss};
\draw (182.83,112.83) node [anchor=north west][inner sep=0.75pt]   [align=left] {S2};
\draw (169,208.4) node [anchor=north west][inner sep=0.75pt]    {$N_{\mathrm{vac}}$};
\draw (147,132.4) node [anchor=north west][inner sep=0.75pt]    {$N_{\mathrm{src}}$};
\draw (234,163) node [anchor=north west][inner sep=0.75pt]   [align=left] {Loss};
\draw (374,93) node [anchor=north west][inner sep=0.75pt]   [align=left] {Loss};
\draw (374,133) node [anchor=north west][inner sep=0.75pt]   [align=left] {Loss};
\draw (374,163) node [anchor=north west][inner sep=0.75pt]   [align=left] {Loss};
\draw (374,193) node [anchor=north west][inner sep=0.75pt]   [align=left] {Loss};
\draw (374,223) node [anchor=north west][inner sep=0.75pt]   [align=left] {Loss};
\draw (147,152.4) node [anchor=north west][inner sep=0.75pt]    {$|0\rangle $};
\draw (147,112.4) node [anchor=north west][inner sep=0.75pt]    {$|0\rangle $};
\draw (146,192.4) node [anchor=north west][inner sep=0.75pt]    {$|0\rangle $};
\draw (147,220.4) node [anchor=north west][inner sep=0.75pt]    {$|0\rangle $};
\draw (131,22) node [anchor=north west][inner sep=0.75pt]   [align=left] {(a) BSBS model};
\draw (131,271) node [anchor=north west][inner sep=0.75pt]   [align=left] {(b) Coherent light model};
\draw (311,400.4) node [anchor=north west][inner sep=0.75pt]    {$U_{in}$};
\draw (168,438.4) node [anchor=north west][inner sep=0.75pt]    {$N_{\mathrm{vac}}$};
\draw (146,422.4) node [anchor=north west][inner sep=0.75pt]    {$|0\rangle $};
\draw (147,450.4) node [anchor=north west][inner sep=0.75pt]    {$|0\rangle $};
\draw (147,292.4) node [anchor=north west][inner sep=0.75pt]    {$|0\rangle $};
\draw (147,322.4) node [anchor=north west][inner sep=0.75pt]    {$|0\rangle $};
\draw (147,361.4) node [anchor=north west][inner sep=0.75pt]    {$|0\rangle $};
\draw (147,392.4) node [anchor=north west][inner sep=0.75pt]    {$|0\rangle $};
\draw (201.83,364) node [anchor=north west][inner sep=0.75pt]   [align=left] {D};
\draw (202.83,393) node [anchor=north west][inner sep=0.75pt]   [align=left] {D};
\draw (168,306.4) node [anchor=north west][inner sep=0.75pt]    {$N_{src}$};
\draw (167,378.4) node [anchor=north west][inner sep=0.75pt]    {$N_{src}$};

\end{tikzpicture}

\caption{{\bf Schematic figure of the digital twin.} The structure is shown for the simulation for $N_{\rm src}=2$ photon sources and $N_{\rm src}+N_{\rm vac}=N_{\rm mod}=4$ interferometer modes. Panel (a) of the figure depicts the programmable boson sampler in Figure~\ref{fig:Fig1} (BSBS setup) with the two-mode squeezing gates S2, interferometer, and loss channels. Panel (b) shows the classical analogous setup using coherent light generated by the displacement gates D.}
\label{fig:model}
    
\end{figure}

Figure~\ref{fig:model} shows the model used for the simulation of the programmable boson sampler, see again Fig.~\ref{fig:Fig1} in main text. From $N_{\rm src}+N_{\rm vac}$ vacuum states as inputs, $N_{\rm src}$ two-mode squeezing gates with squeezing parameters 
\begin{equation}
r_i = \operatorname{asinh}\!\left(\sqrt{\langle n\rangle}\right),
\end{equation}
with the desired average photon number $\langle n\rangle$ per mode, generate correlated pairs of photons. For the evaluations in the work, a value of $\lambda=0.65$ in Eq.~\eqref{photonsource} was used, corresponding to a squeezing parameter $r=0.7753$ and average photon number $\langle n\rangle=0.7316$. For each squeezing gate, one output is heralded, and one output is fed into the interferometer. The action of the squeezing gates, loss channels, and interferometer is efficiently simulated using a Gaussian description using the mean and covariance of its quadratures on $N_{\mathrm{src}}+N_{\mathrm{mod}}$ modes. For the photon-number moments $\langle \hat n_i \hat n_j...\rangle$ in Eq.~\eqref{eq:photonmoments}, numerically exact expectation values are calculated from the Gaussian state. For the comparisons with classical light in Section~\ref{sec:coherent}, the two-mode squeezing gates are replaced by displacement gates to generate coherent light as input for the interferometer. The displacement parameter $\alpha=\sqrt{\langle n'\rangle}$ is chosen such that the average photon number is $\langle n'\rangle=2\langle n\rangle=1.46$ so that the optical power is equal in both cases.


\subsection{Training of photonic reservoir computing model}\label{app:PRCM_training}
The readout weights in the BSBS-RCM are obtained via the Moore–Penrose pseudo-inverse of the state matrix, an estimator that is exact but numerically reliable only when the matrix is well-conditioned. At smaller encoding scales (L96, $N_{\rm mod}=8/12$), the state matrix remains well-conditioned and the pseudo-inverse trains reliably. At the largest configuration (L96, $N_{\rm mod}=16$), the state matrix is rank-deficient by roughly two orders of magnitude, a property of the encoding itself, and the pseudo-inverse becomes numerically unstable, collapsing the trained model's prediction capability. Writing the singular value decomposition of the state matrix as $R_{\text{train}} = U\Sigma V^\top$, this instability is removed by an SVD-based ridge estimator, $W^{\ast}_{\text{out}} = V\,\mathrm{diag}(\sigma_i/(\sigma_i^2+\gamma))\,U^\top Y_{\text{train}}$, where $\sigma_i$ are the singular values of $R_{\text{train}}$ (the diagonal entries of $\Sigma$) and $Y_{\text{train}}$ is the target, with a single regularization strength $\gamma = 10^{-9}$ fixed across the unitary seeds, which is used only for this configuration. The results from smaller configurations are obtained by using the pseudo-inverse.

\section{Derivation of the model for thermal convection flow}\label{app:derivation_convection_flow}
Details of the thermal convection flow model are provided in the following. The flow domain is $A=[0,2\sqrt{2}]\times [0,1]$ in units of the height of the layer $H$ as shown in Fig.~\ref{fig:Fig3}(b). The horizontal coordinate is $x$, the vertical one is $z$. For small temperature differences between bottom and top, heat is carried by diffusion only from the bottom to the top while the fluid is at rest. The resulting temperature profile is linear, $\overline{T}(z) = T_{\rm c}-\frac{\Delta T}{H}z$. This is the equilibrium configuration of the convection flow. The velocity field has two components ${\bm u}(x,z,t)=(u_x(x,z,t), u_z(x,z,t))$. For the following, we consider the temperature deviation $\theta(x,z,t)=T(x,z,t)-\overline{T}(z)$. The temperature deviation and velocity fields, $\theta(x,z,t)$ and ${\bf u}(x,z,t)$, are coupled by the Boussinesq equations \cite{Chilla2012}. They are given by
\begin{align}
{\bm \nabla} \cdot {\bm u}&= 0 \,,
\label{eq:NS1}\\
\frac{\partial {\bm u}}{\partial t} + ({\bm u} \cdot {\bm \nabla}) {\bm u} &= - {\bm \nabla} p + \sigma {\bm \nabla}^2 {\bm u} + {\rm Ra}\, \sigma\, \theta {\bm e}_z \,,\label{eq:NS2b}\\
\frac{\partial \theta}{\partial t} + ({\bm u}\cdot {\bm \nabla})\theta &= {\bm \nabla}^2 \theta + u_z\,
\label{eq:NS3}
\end{align}
where we use the kinematic pressure $p$. The Rayleigh number ${\rm Ra}$ and the Prandtl number $\sigma$ are the two parameters that characterize the strength of the thermal driving via the temperature difference $\Delta T$ and the ratio of momentum to temperature diffusion, respectively.  

The boundary conditions in $x$-direction are periodic. At $z=0,1$, one takes
\begin{align}
u_z\big|_{z=0,1}=0,\quad \frac{\partial u_x}{\partial z}\Bigg|_{z=0,1}=0\quad\mbox{and}\quad \theta\big|_{z=0,1}=0\,.\label{bc1}
\end{align}
They correspond to isothermal, impermeable, free-slip walls. Incompressibility and two-dimensionality allow to reduce the velocity vector field further to a scalar stream function $\zeta(x,z,t)$ by $u_x=-\frac{\partial \zeta}{\partial z}$ and $u_z=\frac{\partial \zeta}{\partial x}$. This ansatz satisfies \eqref{eq:NS1} automatically and the equations of motion \eqref{eq:NS2b}--\eqref{eq:NS3} are now given by two scalar functions only,
\begin{align}
\frac{\partial \nabla^2\zeta}{\partial t}&=\frac{\partial \zeta}{\partial z}  \frac{\partial \nabla^2\zeta}{\partial x}-\frac{\partial \zeta}{\partial x}  \frac{\partial \nabla^2\zeta}{\partial z} + \sigma \nabla^4\zeta +{\rm Ra} \sigma\frac{\partial \theta}{\partial x}\,,\label{2d1}\\   
\frac{\partial \theta}{\partial t}&=\frac{\partial \zeta}{\partial z}  \frac{\partial \theta}{\partial x}-\frac{\partial \zeta}{\partial x}  \frac{\partial \theta}{\partial z} + \nabla^2\theta +\frac{\partial \zeta}{\partial x}\label{2d2}\,,
\end{align}
with boundary conditions in the vertical direction
\begin{align}
\zeta\big|_{z=0,1}=0,\quad \frac{\partial^2\zeta}{\partial z^2}\Bigg|_{z=0,1}=0\quad\mbox{and}\quad \theta\big|_{z=0,1}=0\,.\label{bc}
\end{align}
The nonlinear coupled partial differential equations (PDE) \eqref{2d1} and \eqref{2d2} are then reduced by an expansion into trigonometric Fourier modes, which satisfy the boundary conditions for the stream function and temperature, both in the horizontal and vertical directions. The Fourier modes encode the possible spatial structure, and thus reduce the original system of nonlinear PDEs into an infinite system of nonlinear coupled ordinary differential equations. The {\em finite dimension} of the model (or the finite number of degrees of freedom) follows by a truncation of this set. This is the basic idea of a Galerkin model \cite{Holmes2012}. Here, we take four modes for the expansions of temperature deviation $\theta$ and four modes for the stream function $\zeta$, which leads to eight time-dependent amplitudes $X_0(\tau)$ to $X_7(\tau)$, see also refs. \cite{Pfeffer2022,Rathor2026}.  

The following truncated expansion with a total of eight Fourier modes, all of which satisfy the boundary conditions \eqref{bc}, is used to arrive at the truncated Galerkin model,
\begin{align*}
\zeta(x, z, t)= & C_\zeta [X_0(t) \sin (\alpha x) \sin (\beta z)+X_3(t) \sin (\beta z) +\\
& X_4(t) \cos (\alpha x) \sin (2 \beta z)+X_6(t) \sin (3 \beta z)] \\
\theta(x, z, t)= & C_\theta[\sqrt{2} X_1(t) \cos (\alpha x) \sin (\beta z)-X_2(t) \sin (2 \beta z) +\\
& X_5(t) \sin (\alpha x) \sin (2 \beta z)+X_7(t) \sin (4 \beta z)]
\end{align*}
with the normalization constants
\begin{align*}
	C_\zeta & =\sqrt{2} \frac{\alpha^2+\beta^2}{\alpha \beta} \\
	C_\theta & =\frac{\left(\alpha^2+\beta^2\right)^3}{\alpha^2 \beta \mathrm{Ra}},
\end{align*}
where $\alpha$ and $\beta$ are the horizontal and vertical wavenumbers, respectively, representing the flow in the fluid layer described by the resulting Lorenz-type model. 

The Lorenz-type model equations for 8 degrees of freedom ${X_0, X_1, ..., X_7}$, the L8 model, are given by
\begin{subequations}\label{eq:L8}
    \begin{align}
    \frac{d X_{0}}{d \tau}=\, & \sigma\left(X_{1}-X_{0}\right)-\frac{c_{3}}{\sqrt{2} c_{1}} X_{3} X_{4}+\frac{3 c_{-5}}{\sqrt{2} c_{1}} X_{4} X_{6} \\
    \frac{d X_{1}}{d \tau}= & -X_{1}+r X_{0}+\frac{1}{2} X_{3} X_{5}-\frac{3}{2} X_{5} X_{6}-X_{0} X_{2} \\
    \frac{d X_{2}}{d \tau}= & -b X_{2}+X_{0} X_{1} \\
    \frac{d X_{3}}{d \tau}= & -\frac{\sigma b}{4} X_{3}+\frac{3}{2 \sqrt{2}} X_{0} X_{4} \\
    \frac{d X_{4}}{d \tau}= & -\sigma \frac{c_{4}}{c_{1}} X_{4}-\sigma \frac{c_{1}}{\sqrt{2} c_{4}} X_{5}+\frac{c_{0}}{\sqrt{2} c_{4}} X_{0} X_{3} \nonumber \\
    & -\frac{3 c_{-8}}{c_{4}} X_{0} X_{6} \\
    \frac{d X_{5}}{d \tau}=& -\frac{c_{4}}{c_{1}} X_{5}-\sqrt{2} r X_{4}-X_{3} X_{1}+3 X_{6} X_{1} \nonumber \\ 
    & -2 \sqrt{2} X_{4} X_{7} \\
    \frac{d X_{6}}{d \tau}=& -\frac{9 \sigma b}{4} X_{6}-\frac{1}{2 \sqrt{2}} X_{0} X_{4} \\
    \frac{d X_{7}}{d \tau}= & -4 b X_{7}+\sqrt{2} X_{4} X_{5}.
    \end{align}
\end{subequations}
Here, $c_n = \alpha^2 + n \beta^2$, where $n \in \mathbb{Z}$. The parameters are the Prandtl number $\sigma = 10$, $b=8/3$, and the relative Rayleigh number $r=(Ra-Ra_c)/Ra_c=28$ with the critical Rayleigh number $Ra_c$ for the onset of convection. For these dynamical system parameters, $\alpha=\pi/\sqrt{2}$ and $\beta=\pi$.  The original time $t$ of the fluid flow equations is rescaled to $\tau=c_1t$.

\section{Details of the classical reservoir computing model}\label{app:CRCM}
In classical reservoir computing model, the reservoir state is updated as follows \cite{Rathor2026}
\begin{align*}
	\label{eq:crcm_updatation}
	{\bm r}(t+\Delta t)&=(1-\varepsilon){\bm r}(t)\\ 
                       &+ \varepsilon \tanh[W\bm{r}(t) + W_{\rm in}\bm{X}(t+\Delta t)] \,,
\end{align*}
where the reservoir matrix $W$ is characterised by the spectral radius $\rho$, $W_{\rm in}$ is the input weight matrix, and $\bm{X}$ is the input vector from the prediction task. The output is given by the linear system $\bm{y} = W_{\rm out}\bm{r}$, where $W_{\rm out}$ is determined by ridge regression with parameter $\lambda_{\rm ridge}$. The optimal hyperparameters for the classical RCM are found in  a grid search. This is performed on 11 different seeds of the reservoir matrix, each with the following grid. The spectral radii of the reservoir matrix take values of $\rho \in [0, 1.3]$ increasing in steps of 0.1; the ridge regression parameters (to avoid overfitting) take values of $\lambda_{\rm ridge} \in [10^{-11}, 10^{-2}]$ which are increased in factors of 10; the leaking rate takes values of $\varepsilon \in [0.01, 1]$ with the increment refined systematically. The number of reservoir nodes, $N_r=2048$, and the density of active reservoir nodes, $D=0.08$, is kept fixed. Then the maximum prediction horizon irrespective of $\{\rho, \lambda_{\rm ridge}\}$ is plotted with $\varepsilon$ to identify the optimal leaking rate $\varepsilon = 0.7$ for L8 model and $\varepsilon = 0.02$ for L96 model. Subsequently, three potential pairs $\{\rho, \lambda_{\rm ridge}\}$, based on their number of occurrences, are selected. The robustness of these sets of hyperparameters is tested against 100 seeds for the reservoir matrix. The results for the C-RCM presented in the main text correspond to the hyperparameter set $\{\rho, \lambda_{\rm ridge}, \varepsilon\} = \{0.1, 10^{-5}, 0.7\}$ for L8 model and $\{\rho, \lambda_{\rm ridge}, \varepsilon\} = \{0.6, 10^{-3}, 0.02\}$ for L96 model.

\bibliography{2025_refs_qcml}

\end{document}